\documentclass[12pt]{article}
\usepackage{graphicx}
\usepackage{anysize}
\usepackage{graphics}
\newcommand{\be}{\begin{equation}}
\newcommand{\ee}{\end{equation}}
\newcommand{\bea}{\begin{eqnarray}}
\newcommand{\eea}{\end{eqnarray}}

\def\4vol{{\int d^4x \sqrt{-g}}}

\def\simlt{\stackrel{<}{{}_\sim}}

\newcommand{\nc}{\newcommand}

\nc{\nt}{\tilde{N}}
\nc{\ra}{\rightarrow}
\nc{\lsim}{\begin{array}{c}\,\sim\vspace{-21pt}\\< \end{array}}
\nc{\gsim}{\begin{array}{c}\sim\vspace{-21pt}\\> \end{array}}
\nc{\tnt}{\tilde{N}}
\nc{\tst}{\tilde{t}}

\nc{\LL}{L}
\nc{\vv}{\tilde{v}}

\title{
\vspace*{-1.3cm}
\begin{flushright}
\normalsize{
ANL-HEP-PR-06-54\\
EFI-06-09
}
\end{flushright}
\vspace{0.5cm}
\Large
\textbf{Gravitons and Dark Matter in Universal Extra Dimensions}
\vspace*{1.0cm}
\author{\textbf{Nausheen R.~Shah$^{a,c}$ and Carlos E.M.~Wagner$^{a,b,c}$}\\
\\[1.5cm]
\normalsize\emph{Enrico Fermi Institute$^a$ and Kavli Institute
for Cosmological Physics$^b$,}\\
\normalsize\emph{University of Chicago,
 Chicago, IL 60637, USA} \\
$^c$\normalsize\emph{HEP Division, Argonne National Laboratory,
9700 Cass Ave.,
Argonne, IL 60439, USA}}}

\marginsize{3cm}{3cm}{2cm}{3.7cm}
\begin{document}
\setcounter{page}{0}
\maketitle
\vspace{0.5cm}
\begin{abstract}
Models of Universal Extra Dimensions (UED) at the TeV scale lead to
the presence of Kaluza Klein (KK) excitations of the ordinary
fermions and bosons of the Standard Model that may be observed at
hadron and lepton colliders. A conserved discrete symmetry,
KK-parity, ensures the stability of the lightest KK particle (LKP),
which, if neutral, becomes a good dark matter particle. It has been
recently shown that for a certain range of masses of the LKP a relic
density consistent with the experimentally observed one may  be
obtained. These works, however, ignore the impact of KK graviton
production at early times. Whether the $G^1$ is the LKP or not, the
 $G^n$ tower thus produced can decay to the LKP, and depending on
the reheating temperature, may lead to a modification of the relic
density. In this article, we show that this effect may lead to a
relevant modification of the range of KK masses consistent with the
observed relic density. Additionally, if evidence for UED is
observed experimentally, we find a stringent upper limit on the
reheating temperature depending on the mass of the LKP observed.
\end{abstract}
\thispagestyle{empty}

\newpage

\setcounter{page}{1}
%%%%%%%%%%%%%%%%%%%%%%%%%%%%%%%%%%%%%%%%%%%%%%%%%%%%%%%%%%%%%%%%%%%%%%%%%%%%%%
%\marginsize{3cm}{3cm}{2.5cm}{2.5cm}

\maketitle

\section{Introduction}
Theoretical models constructed to understand physics beyond the
standard model (SM) (most notably string theory) frequently imply
the existence of extra dimensions. It has been a particular
challenge in phenomenology to understand how these extra dimensions
would be realized and manifested in our observable $3+1$ dimensional
world. The number, shape and size of these dimensions, as well as
the particles allowed to propagate in them give rise to several
different models, all having different phenomenological
implications.

We will be considering universal extra dimensions (UED) where the SM
fields propagate in all the extra dimensions. For toroidal
compactification, this would imply a tower of Kaluza-Klein (KK)
particles for every SM particle, each carrying KK number. Momentum
conservation in the extra dimensions implies KK number conservation.
However, the requirement of obtaining the proper SM chiral modes at
low energies leads to constraints on the possible compactification
geometries. An example for $D=5$ is the orbifold $S^1/Z_2$ where the
$Z_2$ projects out half of the zero modes, leaving only SM fields.
Additionally, this breaks translation invariance along the extra
dimensions, so KK number is no longer conserved. A residual
symmetry, KK-parity, is still present and it is sufficient to ensure
the stability of the lightest KK particle (LKP). It also ensures
that KK particles are always produced in pairs, allowing for good
agreement between theory and experiment for small values of the
compactification scale, of the order of a few hundred
GeV~\cite{Antoniadis:1990ew,Appelquist:2000nn, Appelquist:2002wb}.

The stability of the LKP allows for an interesting candidate for
dark matter (DM)~\cite{Servant:2002aq}. The LKP is expected to be
weakly interacting and electrically neutral if it is to be
considered a candidate for dark matter. The usual candidate is
$B^1$, the KK partner of the hypercharge gauge boson. For $m_{\rm
KK}\sim\mathcal{O}~(1)$~TeV, $B^1$ gives excellent agreement with
the observed relic density. There have been many further analyses of
the relic density in UED, including a more proper treatment of
coannihilation effects, and the impact of the inclusion of second KK
level resonances~\cite{Griest:1990kh, Kakizaki:2005en,
Kakizaki:2005uy,Kong:2005hn,Burnell:2005hm,Kakizaki:2006dz}.

However, most of these studies ignore the gravitational sector. We
shall mostly work within the context of one extra dimension. The
gravitons couple extremely weakly and therefore are generally
considered to be unimportant for collider studies. Cosmologically,
however, they can have a significant effect~\footnote{The radion can
also be quite relevant~\cite{Kolb:2003mm}. We shall assume that the
radion is inert during inflation and that it acquires a large mass,
and therefore it does not have any impact on our analysis.}, and
they can also be a candidate for the LKP~\cite{Feng:2003nr}.

We are interested in investigating the inclusion of the graviton
tower in the scenario with $B^1$ as the LKP. The gravitons do not
evolve thermally, and have very long lifetimes, so if they are
present, we expect them to decay to the LKP sometime after Big Bang
Nucleosynthesis (BBN). The preservation of the light element
abundances sets a bound on the amount of energy that can be released
in a decay~\cite{Ellis:2005ii,Kawasaki:2004qu,Cyburt:2002uv}.
Secondly, since the gravitons decay ultimately into the LKP, we
expect the relic density to increase. Therefore, the mass of the LKP
consistent with the observed relic density is lowered. Additionally,
if any of the gravitons decay after matter domination, we have to
consider the effects of non-thermalized photons released in the
decays on the spectrum of the diffuse photon flux
(DPF)~\cite{Sreekumar:1997yg}.

This article is organized as follows. In Section 2, we review the
standard calculation for relic density. Section 3 reiterates the
calculation for the density of gravitons presented in
Ref.~\cite{Feng:2003nr}, emphasizing the points relevant for our
analysis. In Section 4, we analyze the lifetimes for the decay of
the graviton tower and study its implication on the diffuse photon
spectrum. Section 5 goes over the constraints on the energy released
in the decay of the KK gravitons to the LKP. In Section 6, we
combine the above results with the standard relic density
calculation. Our goal is to analyze the effect of the gravitons on
the predicted mass of the LKP consistent with the known dark matter
density. We will show that, in fact, almost any value of $m_{\rm
KK}$ lower than the one obtained in the absence of gravitons would
be allowed, provided the reheating temperature, $T_R$, is large
enough. For the calculation of the relic density of $B^1$, we ignore
complicating factors such as coannihilation and second KK level
resonance effects~\cite{Griest:1990kh, Kakizaki:2005en,
Kakizaki:2005uy,Kong:2005hn,Kakizaki:2006dz}. However, since these
effects can be parameterized in the effective cross-section, we
present the effect the graviton tower will have on the mass of the
LKP without the gravitons, $m_{\rm WG}$, and note that as $m_{\rm
WG}$ increases, the contribution due to the graviton tower becomes
large, and should be included in any precise calculation of $m_{\rm
KK}$. We also find that all the gravitons, except for $G^1$, decay
right after BBN. At these early times, the electromagnetic BBN
constraints are very weak. However it has been shown that at small
lifetimes, the hadronic constraints are very
stringent~\cite{Kawasaki:2004yh, Jedamzik:2004er}. Therefore,
including constraints from both hadronic and electromagnetic decays,
requiring consistency with BBN light element predictions, we find a
stringent limit on the mass difference between $m_{G^1}$ and the LKP
mass, $m_{\rm KK}$. Since the $G^1$ is long-lived, we also derive
another constraint on the mass difference such that the observed
spectrum of the DPF is not destroyed. Comparing the two constraints,
we find that there exists a region of parameter space where both
constraints are satisfied and which gives mass differences of the
same order of magnitude as that obtained by radiative loop
corrections to $m_{B^1}$~\cite{Cheng:2002iz}. Additionally, we
observe that if experimental evidence for UED is found, and the
relic density induced by standard interactions of the KK modes is
found to be lower than the one observed experimentally, the
reheating temperature may be determined by assuming that the
graviton KK modes provide the contribution necessary to achieve
consistency between theory and experiment. Alternatively, including
the possibility of other unobserved exotic particles contributing to
the relic density, an upper bound on the reheating temperature is
obtained.

\section{Relic Calculation}

Let us start by recalling the computation of the density of thermal
relics. For any particle $z$, the evolution of the number density
$n_z$ is
governed by the Boltzmann equation:

\begin{eqnarray}
\label{eq:Boltz}
\frac{dn_z}{dt}+3Hn_z &=& -\left<\sigma v\right>(n_z^2-n_{eq}^2) \\
\label{eq:H}
                H &=& \left(\frac{8\pi G_N\rho}{3}\right)^{1/2}\\
\label{eq:neq}
           n_{eq} &=& g\left(\frac{mT}{2\pi}\right)^{3/2}e^{-m/T}
\end{eqnarray}
where $\left<\sigma v\right>$ is the thermally averaged
cross-section multiplied by the relative velocity, and
Eq.~(\ref{eq:neq}) is valid in the non-relativistic approximation,
with $m$ and $g$ being the mass and number of degrees of freedom
associated with the particle $z$.
The temperature at which the particle decouples from the thermal
bath is denoted by $T_F$ (\textit{freeze-out temperature}) and
roughly corresponds to when $\Gamma=n\left<\sigma v\right>$ is of
the same order as $H$.

Changing variables from $n$ to $Y=n/s$ (we will drop the subscript $z$ from
now on), where the entropy density is
given by $s=\frac{2\pi^2}{45}g_*T^3$, and using the fact that $sR^3$
remains constant, we obtain,

\begin{eqnarray}
\label{eq:Y} \dot{Y}s &= & -\left<\sigma v \right>s^2(Y^2-Y_{eq}^2).
         \end{eqnarray}
Introducing the variable $x=\frac{m}{T}$, in the radiation dominated
era:
\begin{eqnarray}
\label{eq:Hrad} H^2&=& \frac{\pi^2g_*T^4}{180M^2_{4}} \; ,
 \;\;\;\;\;\;\;\;\;\;\;
  t=\frac{1}{2H}
\;\;\;\;\;\;\;\;\;\;
\; \Rightarrow \; \frac{dx}{dt}= H \; x.
\end{eqnarray}
Then, Eq.~(\ref{eq:Y}) may be rewritten as
\begin{equation}
\label{eq:Yx} \frac{dY}{dx}=-\frac{\left<\sigma
v\right>}{Hx}s\left(Y^2-Y_{eq}^2\right).
\end{equation}

We are interested in obtaining the relic density, or the equivalent
$Y_\infty$, at late times. The solution of Eq.~(\ref{eq:Yx}) enables
the determination of the relic density and also the value of $x_F$
as a function of $\Gamma$. The ratio of the mass to the freeze-out
temperature, $x_F=m_{\rm KK}/T_F$ is approximately given by
\begin{equation}
\label{eq:xf}
x_F=\ln\left(c(c+2)\sqrt{90\pi}\frac{g}{2\pi^3}\frac{m_{\rm
KK}M_{4}\left<\sigma v\right>}{g_*^{1/2}x_F^{1/2}}\right)
\end{equation}
and can be solved iteratively, with $c \simeq 1/2$ and
$M_{4}=1.7\times10^{15}$ TeV. As calculated in
\cite{Servant:2002aq}, for $B^1$ as the LKP, without including
co-annihilation~\footnote{The effects of including these corrections
to the cross-section in our analysis are discussed in detail in
Section 6.3.}, the non-relativistic approximation is:
\begin{eqnarray}\label{eq:sigma_r}
\left<\sigma\nu\right>&=&a+\frac{6b}{x}\\
&&\nonumber\\
          a&=&\frac{4\pi\alpha_1^2\left(2Y_f+3Y_B\right)}{9m_{\rm KK}^2}=
\frac{a'}{m_{\rm KK}^2} \; ,\\
          b&=&-\frac{\pi\alpha_1^2\left(2Y_f+3Y_B\right)}{18m_{\rm KK}^2}=
\frac{b'}{m_{\rm KK}^2} \; ,
\end{eqnarray}
where $Y_f$ and $Y_B$  denote a summation over the fourth power of
the hypercharges of fermions and bosons produced in the annihilation
of $B^1$ respectively. Moreover, ignoring the splitting between the
different KK states of a given tower, one obtains, approximately
\begin{equation}\label{eq:xf2}
x_F=\ln\left[\frac{1.79\times10^{14}}{x_F^{1/2}}\frac{\mbox{
TeV}}{m_{\rm KK}}\left(a'+\frac{6b'}{x_F}\right)\right].
\end{equation}

In terms of this, $Y_\infty$ is found to be:
\begin{eqnarray}
\label{eq:Y_infty}
Y_{\infty}^{-1}&=&\frac{4\pi}{3}\sqrt{\frac{ g_*}{5}}M_4m_{\rm KK}x_F^{-1}(a+\frac{3b}{x_F})\nonumber \\
&&\nonumber\\
   &=&\frac{3.05\times10^{16}}{x_F}\frac{\mbox{TeV}}{m_{\rm KK}}(a'+\frac{3b'}{x_F})
\end{eqnarray}
where we have assumed masses of $\mathcal{O}~(1)$ TeV, and used $g_*
\simeq 92$ for $T_F\sim50$ GeV and $s_0=2889.2 \mbox{ cm}^{-3}$. For
more than one extra dimension this analysis is more complicated but
a rough estimate may be obtained by assuming independent towers of
KK modes for each extra dimension, $d$. The current relic energy
density then is simply given by $\rho=m_{\rm KK} \; s_0 \; d
\;Y_{\infty}$. We will however restrict ourselves to one extra
dimension in most of our numerical work.

As emphasized in the introduction, the above derivation for the
density of $B^1$ ignores the contribution of the KK modes of the
gravitons, which may become relevant for sufficiently large values
of the reheating temperature $T_R$.

\section{Density of KK Gravitons}
After reheating, except for gravitons, all other particles are
initially in equilibrium and follow the pattern described in the
previous section. After dilution of their density due to inflation,
the gravitons are produced mainly in collisions involving gauge
bosons, either in the initial or the final state, with
cross-sections proportional to $\alpha_i/M_4^2$, where $\alpha_i$ is
the relevant gauge coupling. Since their density is so small, they
are never in equilibrium and further we can ignore graviton
annihilation processes to first order. Because the decay times for
gravitons are of the order of $\tau\sim 10^7$ s, the gravitons
produced in the early universe are still present when the LKP
freezes out. As the universe cools down further, the gravitons start
to decay. Including higher order mass
corrections~\cite{Cheng:2002iz}, we will have decays of the form:

\begin{equation}\label{eq:Decay}
G^n\rightarrow n \; \mbox{LKP}+X \nonumber
\end{equation}
where $X$ denotes SM particles. Here we are considering only KK
number preserving decays since the KK number violating decays would
only take place at the orbifold fixed points and so would only have
a sizeable contribution in the case when the KK number preserving
decays are suppressed due to phase space considerations. Hence,
after decaying, the KK gravitons contribute to the LKP relic density
and a proper computation of the dark matter density demands the
addition of this effect to the relic density calculated in the
previous section.

In order to obtain a quantitative estimate of this effect, we follow
the derivation for the number density of gravitons presented in
Ref.~\cite{Feng:2003nr}. The number density of the KK gravitons at
each level $n$ is again determined by the Boltzmann equation:
\begin{equation}
\label{eq:Boltz_G} \frac{dn_{G^n}}{dt}+3Hn_{G^n}=C_{G^n},
\end{equation}
where $C_{G^n}$ is the collision operator and can be parameterized
as follows:
\begin{equation}
\label{eq:CG} C_{G^n}=C\sigma[g_*(T)n_0]^2,
\end{equation}
where
\begin{equation}
\label{eq:sigma_G} \sigma=\frac{\alpha_3}{4\pi M^2_4}.
\end{equation}
Here $\alpha_3$ is the strong coupling constant and $C$ can be
understood as the fraction of all possible collisions which will
interact strongly to produce gravitons. The exact calculation of the
total production cross section is quite complicated, but can be
estimated by the method presented in Appendix~\ref{appendixA}. The
authors of Ref.~\cite{Feng:2003nr} used an analogy with the
calculation of gravitino abundances~\cite{Bolz:1998ek, Bolz:2000fu},
and estimated $C\sim\mathcal{O}~(0.01)$. However the sample
calculation of the cross-section for the production of relativistic
gravitons presented in Appendix~\ref{appendixA} leads us instead to
expect $C\sim\mathcal{O}~(1)$. For completeness we will consider a
range of values, $0.01\leq C\leq1$, in our numerical work.

For UED theories, $g_*(T)=g_*^{KK}D_d(T)$, where $g_*^{KK}$ is the
effective number of degrees of freedom per KK level. $D_d(T)$ can be
approximated by counting all modes with masses below $T$:

\begin{equation}\label{eq:D_d}
D_d(T)=\frac{1}{2^d}V_d\left[\frac{T}{m_{\rm KK}}\right]^d
\end{equation}
where
\begin{equation}\label{eq:Vol}
V_d=\frac{\pi^{d/2}}{\Gamma(1+\frac{d}{2})}=2,\pi,\frac{4}{3}\pi,
\frac{1}{2}\pi^2,... \;\;\; \mbox{ for d} =1,2,3...\;,
\end{equation}
is the $d$ dimensional volume of a unit sphere, and the factor
$1/2^d$ in Eq.~(\ref{eq:D_d}) accounts for the restriction to
non-negative $n$. Assuming entropy conservation, $S=sR^3\propto
g_*(T)T^3R^3\propto T^{3+d}R^3$:

\begin{eqnarray}
\label{eq:ent_cons}
\frac{1}{s}\frac{ds}{dt}=-3\frac{1}{R}\frac{dR}{dt}=-3H,&&\frac{dT}{dt}=-\frac{3}{3+d}HT.
\end{eqnarray}

 As in the previous section, making
the substitution $Y=n/s$ and with the help of
Eq.~(\ref{eq:ent_cons}) the Boltzmann equation becomes:
\begin{equation}\label{eq:YGndiff}
\frac{dY_{G^n}}{dT}=-\frac{3+d}{3}\frac{1}{HTs}C\sigma[g_*(T)n_0]^2.
\end{equation}

$Y_{G^n}$ changes until $G^n$ production stops at temperatures
$T\sim nm_{\rm KK}$ and then remains constant until gravitons begin
to decay. After BBN and before gravitons decay:

\begin{equation}\label{eq:YGn}
\begin{array}{cclc}
Y_{G^n}&=& \displaystyle
\frac{45\sqrt{5}\zeta^2(3)}{2\pi^8}\alpha_3\frac{m_{\rm
KK}}{M_{4}}C\sqrt{g_*^{KK}}\frac{3+d}{2+d}\sqrt{\frac{V_d}{2^d}}\left[\left(\frac{T_R}{m_{\rm
KK}}\right)^{1+\frac{d}{2}}-n^{1+\frac{d}{2}}\right]& \mbox{for
}n<\displaystyle\frac{T_R}{m_{\rm KK}};\nonumber\\
\\
 &=&0& \mbox{for }n>\displaystyle\frac{T_R}{m_{\rm KK}}.\nonumber\\
\\
\end{array}
\end{equation}
To find the $Y_G$ contribution to the LKP, assuming only KK number
preserving decays,
\begin{eqnarray}\label{eq:YG}
Y_G&=&\int_0^{T_R/m_{\rm KK}}nY_{G^n}\,d^dn\nonumber\\
   &=&\frac{45\sqrt{5}\zeta^2(3)}{2\pi^8}\alpha_3\frac{m_{\rm KK}}{M_{4}}C\sqrt{g_*^{KK}}\frac{3+d}{(1+d)(4+3d)}\sqrt{\frac{V_dA_d^2}{2^{3d}}}\left[\frac{T_R}{m_{\rm KK}}\right]^{2+\frac{3d}{2}}\nonumber\\
   &=&\alpha(d)\;C\;\frac{m_{\rm KK}}{\mbox{TeV}}\;\left[\frac{T_R}{m_{\rm KK}}\right]^{2+\frac{3d}{2}}\\
A_d&=&\frac{2\pi^{d/2}}{\Gamma(d/2)}=2,2\pi,... \;\;\; \mbox{for d}=
1,2,...\; ,
\end{eqnarray}
where $\alpha_3\sim 0.1$ and $g_*^{KK}\sim 200$.

\section{Decay Lifetimes and Diffuse Photon Flux}

The decay widths for the decay of the gravitons are calculated in
Appendix~\ref{appendixB}. As shown there, the decay of $G^n$ for
$n>1$ is primarily into gauge bosons of KK number $n/2$ ($(n \pm
1)/2$ for $n$ odd), assuming that all fermions are heavier. The
lifetime for these is suppressed by powers of $n$, and is given by:

\begin{eqnarray}
\tau(G^n)&=&\frac{32\pi}{\sqrt{2}\cos^2\theta_W}\frac{M_4^2}{m_n^3}\sqrt{\frac{m_n}{\Delta_n}}\nonumber\\
&\sim&1.76\times10^5\mbox{
s}\left[\frac{\mbox{TeV}}{m_n}\right]^{5/2}\left[\frac{\mbox{TeV}}{\Delta_n}\right]^{1/2}\nonumber \\
&\sim& 5.56\times10^6 \mbox{
s}\frac{1}{n^3}\left[\frac{\mbox{TeV}}{m_{\rm
KK}}\right]^3\label{eq:Gn_life}
\end{eqnarray}
where $\Delta_n \equiv m_{G^n}-2m_{B^{\frac{n}{2}}}<<m_{G^n}$ and in
the last line we have approximated $\Delta_n \sim n\Delta_1\equiv
n(m_{G^1}-m_{B^1})$ with $\Delta_1\sim 10^{-3}m_{\rm KK}$, which is
of the order of the mass corrections induced at one
loop~\cite{Cheng:2002iz}, and, as we will show below, is also of the
order of the mass differences required to satisfy the
phenomenological constraints coming from BBN and diffuse gamma ray
constraints.

The only long-lived graviton is $G^1$, with a lifetime given by:
\begin{eqnarray}
\tau(G^1)&=&\frac{3\pi}{\cos^2\theta_W}\frac{M_4^2}{\Delta_1^3}\nonumber\\
&\sim&2.33\times10^4\mbox{ s}\left[\frac{\mbox{TeV}}{\Delta_1}\right]^3,\nonumber\\
&\sim& 2.33\times 10^{13}\mbox{ s}\left[\frac{\mbox{TeV}}{m_{\rm
KK}}\right]^3,\label{eq:G1_life}
\end{eqnarray}
where we have again assumed in the last line that
$\Delta_1 \sim 10^{-3} m_{\rm KK}$. The different dependence on
the mass difference in Eq.~(\ref{eq:Gn_life}) and (\ref{eq:G1_life})
is due to the fact that one of the decay products for $G^1$ is
massless.

\begin{figure}[!tb]
\centering \scalebox{0.9}[0.8]{\includegraphics[bb=2.9cm 0cm 18.1cm
16.1cm, clip=false]{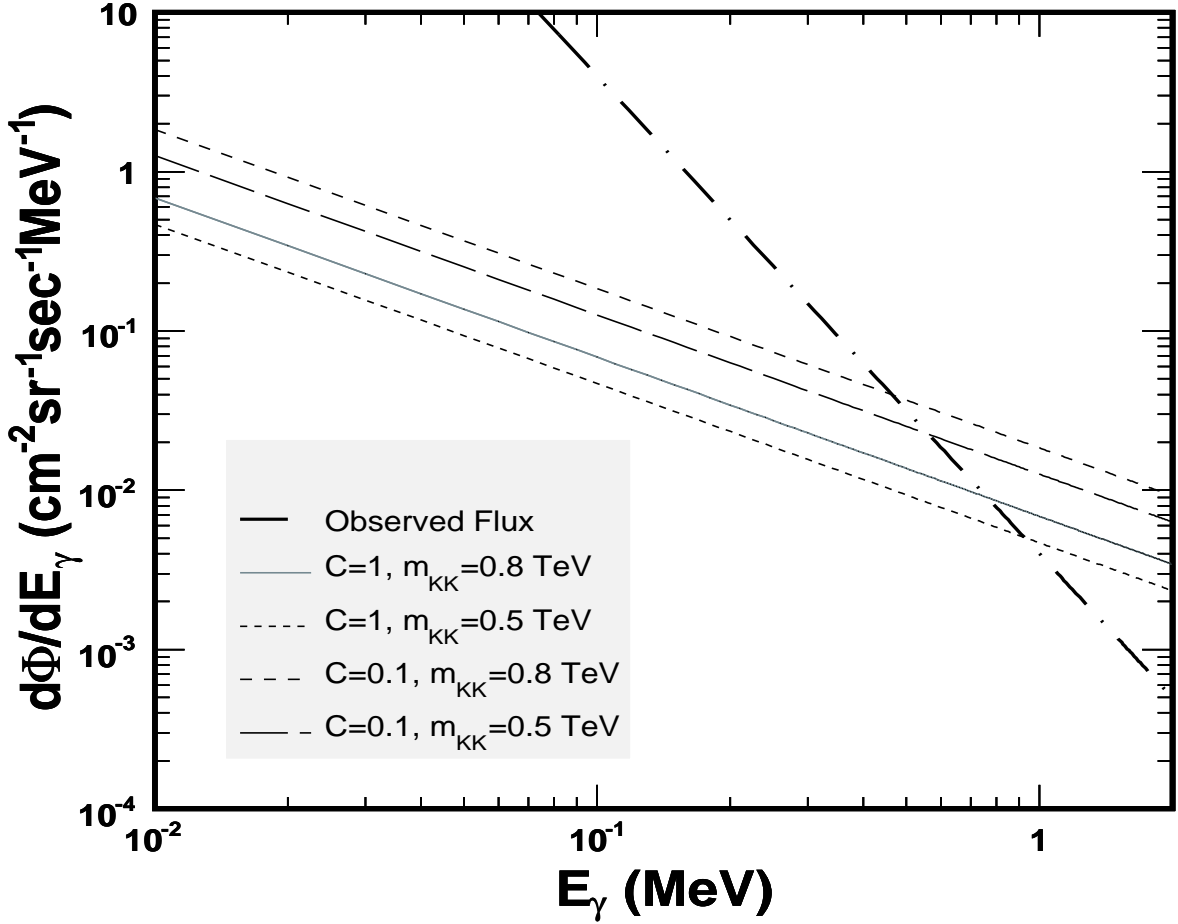}} \caption{\footnotesize{The observed
Diffuse Photon Flux plotted in addition to the maximum flux from
Eq.~(\ref{eq:maxFlux}) for different values of $C$ and $m_{\rm KK}$,
assuming $T_R$ such that $\Omega_{B^1}=0.23$.}}\label{fig:DPF}
\end{figure}

The diffuse photon flux is only sensitive to photons released after
matter domination~\cite{Sreekumar:1997yg,Feng:2003xh} since the late
produced photons don't have time to thermalize. Therefore we just
need to ensure that the energy released in the decay of the $G^1$
does not destroy the spectrum. The diffuse photon flux is given
by~\cite{Feng:2003xh}:

\begin{equation}\label{eq:dPhi/dE}
\frac{d\Phi}{dE_\gamma}\sim\frac{3c}{8\pi}\frac{N^{in}_{G^1}}{V_0\epsilon_\gamma}\left[\frac{t_0}{\tau_{G^1}}\right]\left[\frac{E_\gamma}{\epsilon_\gamma}\right]^{1/2}e^{-\left[\left(\frac{E_\gamma}{\epsilon_\gamma}\right)^{3/2}\frac{t_0}{\tau_{G^1}}\right]}\Theta(\epsilon_\gamma-E_\gamma),
\end{equation}
where $\epsilon_\gamma$ is initial energy released in the decay,
$E_\gamma$ is energy of photons observed now, $t_0$ and $V_0$ are
the current time and volume of the universe, and $N^{in}_{G^1}$ is
the total number of $G^1$s initially. Using $Y_{G^1}$ calculated in
the previous section:

\begin{eqnarray}
\frac{d\Phi}{dE_\gamma}&=&1.63\times10^{-9}\mbox{sr}^{-1}\mbox{s}^{-1}\mbox{cm}^{-2}\mbox{ MeV}^{-1}\left[\left(\frac{T_R}{m_{\rm KK}}\right)^{3/2}-1\right]\nonumber\\
&&\left[\frac{\Delta_1}{\mbox{MeV}}\right]^{3/2}\left[\frac{m_{\rm
KK}}{\mbox{TeV}}\right]\left[\frac{E_\gamma}{\mbox{MeV}}\right]^{1/2}e^{-\left[\left(\frac{E_\gamma\Delta_1}{\mbox{\tiny
MeV}^2}\right)^{3/2}1.85\times10^{-5}\right]}.\label{eq:dPhi/dE2}
\end{eqnarray}

This has maximal value at:
\begin{equation}\label{eq:Emax}
E^{max}_\gamma= \epsilon_\gamma \left(\frac{\tau_{G^1}}{3 t_0} \right)^{2/3}
\simeq
0.687\;\mbox{MeV}\left[\frac{\mbox{GeV}}{\Delta_1}\right].
\end{equation}
Observe that, although $\epsilon_\gamma$ is proportional to
$\Delta_1$, the inverse dependence of $E^{max}_\gamma$ on $\Delta_1$
comes from the energy redshift induced by the expansion of the
Universe ($\tau_{G^1} \propto \Delta_1^{-3}$). This, in turn, leads
to a maximum value for the flux,
\begin{equation}\label{eq:maxFlux}
\frac{d\Phi^{max}}{dE_\gamma}=3.06\times10^{-5}\;\mbox{sr}^{-1}\;\mbox{s}^{-1}\;\mbox{cm}^{-2}\;\mbox{MeV}^{-1}\left[\left(\frac{T_R}{m_{\rm
KK}}\right)^{3/2}-1\right]\left[\frac{\Delta_1}{\mbox{GeV}}\right]\left[\frac{m_{\rm
KK}}{\mbox{TeV}}\right].
\end{equation}
If, for example, we assume a reheating temperature $T_R\sim50 \;m_{\rm KK}$,
we obtain
\begin{equation}\label{eq:dPhi_dEmax}
\frac{d\Phi^{max}}{dE_\gamma}=1.1\times
10^{-2}\;\mbox{sr}^{-1}\;\mbox{s}^{-1}\;\mbox{cm}^{-2}\;\mbox{MeV}^{-1}\left[\frac{\Delta_1}{\mbox{GeV}}\right]\left[\frac{m_{\rm
KK}}{\mbox{TeV}}\right].
\end{equation}
This is two orders of magnitude less than was obtained in
Ref.~\cite{Feng:2003xh} for a similar mass range, and is marginally
consistent with the observed diffuse photon
flux~\cite{Sreekumar:1997yg}. The source of the difference is not
computational, but based on the different assumptions made in both
works. In Ref.~\cite{Feng:2003xh}, the NLKP was assumed to be $B^1$
and had a density comparable to the dark matter relic density, while
in our framework we are concerned with the decay of the $G^1$s,
which have a density much smaller than that associated with the dark
matter relic density.

In order to understand the constraints imposed by the photon flux on
the $G^1$ decay, we can parameterize the experimentally observed
flux in terms of $E_{\gamma}$ as follows,
\begin{equation}
\frac{d\Phi}{dE_\gamma} \simeq
4\times10^{-3}\;\mbox{sr}^{-1}\;\mbox{s}^{-1}\;\mbox{cm}^{-2}\;\mbox{MeV}^{-1}\left[\frac{\mbox{MeV}}{E_\gamma}\right]^3.
\label{eq:obsFlux}
\end{equation}
This is plotted in Fig.~\ref{fig:DPF}. For comparison we have also
plotted the maximum flux according to Eq.~(\ref{eq:maxFlux}) for a
range of values of $C$ and $m_{\rm KK}$, assuming a $T_R$ such that
$\Omega_{B^1}=0.23$. The mass difference allowed by the diffuse
photon flux can then be quantified by requiring that the maximum
differential flux calculated from Eq.~(\ref{eq:maxFlux}) is less
than the observed flux Eq.~(\ref{eq:obsFlux}). It should also be
noted that here we assumed that all the photons produced in the
decay of the gravitons did not thermalize and are available to
distort the spectrum now. This is certainly an overestimate. The
constraints on $\Delta_1$ will be discussed in more detail in
Section~6.6.

\section{Energy Released in Decays}

The electromagnetic showers produced by the decay of $G^n\to n\mbox{
LKP}+X$ can create and destroy light elements. As discussed in
\cite{Ellis:2005ii,Kawasaki:2004qu,Cyburt:2002uv}, this sets
constraints on the lifetime $\tau$ of the unstable $G^n$ as well as
on $\xi_\gamma$, the energy released per background photon in the
decay.

The dependence on $\tau$ can be understood by looking at the
characteristic energy scales in the initially produced photon
spectrum. The primary photon created in the decay interacts with the
background and creates an EM cascade. The fastest interactions are
pair production and inverse compton scattering. These processes
rapidly redistribute the energy and the non-thermal photons reach a
quasi-static equilibrium (QSE). The zeroth order QSE photon spectrum
depends inversely on the temperature of the background plasma. If we
make the \emph{uniform decay} approximation, i.e. all particles
decay at $t=\tau$, this corresponds to a cut-off energy of
$E_C=103\mbox{ MeV }(\tau/10^8\mbox{
s})^{1/2}$~\cite{Cyburt:2002uv}. Therefore higher energy
photo-erosion processes occur for longer lifetime values.

We can understand the $\xi_\gamma$ dependance of photodestruction
and secondary production in a similar way. In the limit of small
$\xi_\gamma$, the decaying particle does not influence the light
element abundances. Beyond this trivial case, we can again use the
uniform decay approximation to gain some insights. As long as a
reaction can take place ($E_{TH}$, the threshold energy for a
reaction, $\simlt E_C$), a typical shower photon has energy:

\begin{equation}\label{eq:PhEn}
\left<E\right>=56\mbox{ MeV }\left(\frac{E_{TH}}{10\mbox{ MeV
}}\right)^{1/2}\left(\frac{\tau}{10^8\mbox{ s}}\right)^{1/4}.
\end{equation}
Therefore, the number of such photons per decay is $N_\gamma\sim
\delta m_{\rm KK}/\left<E\right>$, where $\delta m_{\rm KK}$ is the
energy released in each decay. Thus, the non-thermal photon density
is:

\begin{equation}\label{eq:nTdensity}
n_\gamma=N_\gamma n_{NLKP}=\frac{\xi_\gamma
n_\gamma^{BG}}{\left<E\right>}
\end{equation}

These photons then further thermalize as well as cause
photodestruction of particles to yield other species of particles.
Therefore the dependence of the fractional change in the abundance
of a particle is proportional to $\xi_\gamma$ and inversely
proportional to $\left<E\right>$ and therefore to $\tau$.

Detailed numerical work gives a complicated but weak constraint on
$\xi_\gamma$ for getting the correct light element abundances for
small lifetimes. However after about $\tau>10^7$ s, the dependance
on $\xi_\gamma$ steadies out to $\xi<10^{-15}$ TeV
\cite{Ellis:2005ii,Kawasaki:2004qu,Cyburt:2002uv}. Comparing
Eq.~(\ref{eq:Gn_life}) and (\ref{eq:G1_life}), we see that the only
long-lived graviton is the $G^1$ with $\tau(G^1)\sim 10^{13}\mbox{
s}\left[\frac{\rm TeV}{m_{\rm KK}}\right]^3$, much larger than
$10^7$ s. Hence, electromagnetic energy release bounds only
constrain the decay of the $G^1$ to the LKP.

Similarly, hadrons produced in the decays can scatter off photons,
electrons and, more importantly, background nuclei. Scattering off
photons and electrons just causes them to lose energy, but the
nuclei become more energetic and cause ``hadronic showers''.
Additionally if inelastic scattering occurs, the background nuclei
get dissociated and the light element abundances are changed. Due to
the multitude of possible interactions, these are a lot more
complicated to analyze than the purely electromagnetic case.
Previously it had always been assumed that since the branching ratio
into hadrons is small, the effect would also be negligible. However
detailed numerical work \cite{Kawasaki:2004yh, Jedamzik:2004er}
shows that in fact at early times, the constraints due to hadronic
processes are much stronger then those due to electromagnetic
processes.

As can be seen from Eq.~(\ref{eq:Gn_life}) the lifetime of the
$G^n$s are suppressed by powers of $n$: $\tau(G^n)\sim10^6\mbox{ s
}\frac{1}{n^{3}}\left[\frac{\rm TeV}{m_{\rm KK}}\right]^3. $
Therefore all the $G^n$s except for $G^1$ decay right after BBN. The
bounds at these relatively early decay times depend strongly on the
branching ratio of the decay of the gravitons into hadrons. If all
the visible energy released is in the form of hadrons, the bounds
are quite severe: $\xi_H < 10^{-15}$ TeV. If most of the decay is
into leptons, then the bounds on $\xi_H$ may be significantly
weaker. A preferential decay into leptons could happen if, for any
given tower, the KK leptons are very close in mass to the KK
hypercharge gauge bosons, while the KK quarks are heavier (see
Appendix~\ref{appendixB} for calculation of the graviton decay
widths into fermions). This would be the result one would obtain if
including only the radiative corrections to the KK particle masses
computed in Ref.~\cite{Cheng:2002iz}. Since this is very model
dependent, we will again use the most constrained case of
$\xi_H<10^{-15}$ TeV.

Hence, both for early and late time decays we will use a
conservative bound for the energy released per background photon,
$\xi_B<10^{-15}$ TeV, in our numerical work. Observe that while
strong violations of this bound will certainly induce strong effects
on the light element abundances, and therefore are ruled out, small
violations of this bound are still possible due to the fact that we
used very conservative limits on the energy released.

One loop effects introduce corrections of two forms, constant
(independent of $n$) and proportional to $n$ (bulk and boundary
correction terms)\cite{Cheng:2002iz}. Therefore, the mass of a
particle at KK level $n$ can be written as:

\begin{equation}\label{eq:mass}
m_n=\frac{n}{R}\left(1+\delta\right) +\delta'
\end{equation}
where the corrections $\delta$ and $\delta'$ for each particle type
is the same at each level. The graviton KK modes, of course, do not
receive corrections of this kind at the one loop level, and we shall
assume, for simplicity, that their masses remain unperturbed:
$m_{G^n}=n/R$.

To analyze the effects of the decays, we are going to treat the two
cases ($\delta$, $\delta'=0$) separately. Note that this is only for
ease of understanding. In fact, the true effect would just be the
sum of the two terms. In the decay $G^n\to nB^1+X$ assuming that $X$
is massless to first order, $m_{G^1}\sim m_{B^1}$, and small
momenta, the energy released in each decay is given by:

\begin{equation}\label{eq:Erel}
E_X^n=\frac{m^2_{G^n}-n^2m^2_{B^1}}{2m_{G^n}}\sim
\left\{\begin{array}{ll}
                                            nm_{\rm KK}\delta & \mbox{for }\delta'=0\\
                                            \delta' &\mbox{for }\delta=0.\\
                                            \end{array}\right.\
\end{equation}

For this energy release to be consistent with the light element
abundances \cite{Ellis:2005ii,Kawasaki:2004qu,Cyburt:2002uv} we need
to have:

\begin{equation}\label{eq:xi_bound}
\xi=\int^{T_R/m_{\rm KK}}_0B_{\rm EM/Had}
\frac{n_{G^n}}{n_\gamma}E_X^nd^dn<\xi_B=10^{-15}\mbox{ TeV}
\end{equation}
here $n_{G^n}=s_0Y_{G^n}$ as given in Eq.~(\ref{eq:YGn}), and
$B_{\rm EM/Had}$ is the electromagnetic/hadronic branching ratio.
This fraction depends on whether the LKP is a lepton or a gauge
boson and the mass splittings at each KK level~\cite{Feng:2003uy}.
Throughout this article we choose $B_{\rm EM/Had}=1$ as the most
stringent constraint possible.

\section{Determination of the Lightest KK Particle Mass}

In this section, we will discuss the determination of the lightest
KK particle mass consistent with the observed dark matter relic
density. Since the energy density of LKP is determined, in part, by
the primordial graviton KK density, the lightest LKP mass $m_{\rm
KK}$ will depend on the reheating temperature and the parameter $C$
governing graviton production at early times.

\subsection{$T_R$ Reheating Temperature}
\begin{figure}[!tb]
\centering \scalebox{0.9}[0.8]{\includegraphics[bb=2.9cm 0cm 18.1cm
16.1cm, clip=false]{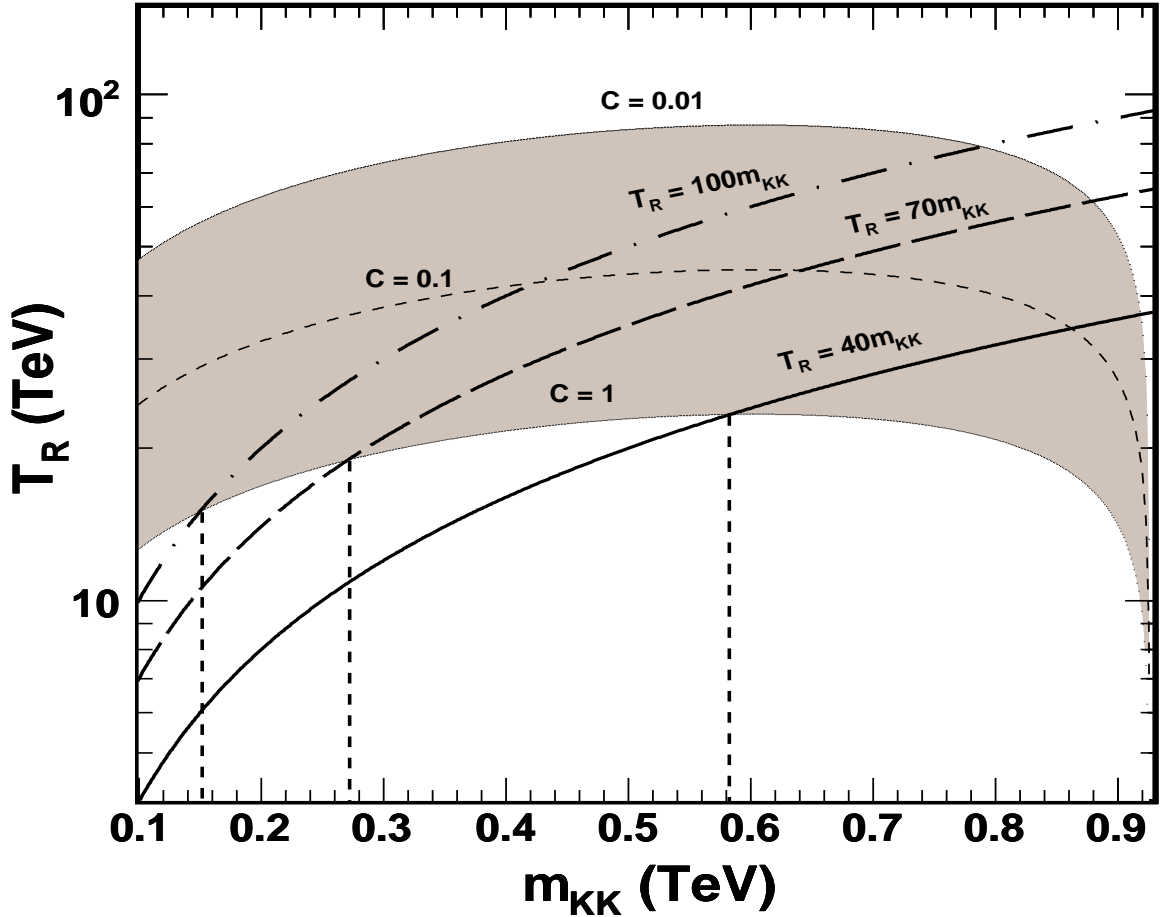}} \caption{\footnotesize{Values of
the reheating temperature $T_R$ obtained by demanding a proper dark
matter density, $\Omega_{B^1}=\Omega_{\rm DM} \simeq 0.23$, for
different values of the graviton production parameter $C=1$, $0.1$,
$0.01$, and assuming $D=5$. Also shown are lines of constant ratios
of the reheating temperature to the lightest KK mass, $T_R=40 \;
m_{\rm KK}$, $70 \; m_{\rm KK}$ and $100 \; m_{\rm KK}$.}
}\label{fig:TrD5}
\end{figure}

Including the decay of the gravitons in the density for $B^1$, we
have:

\begin{figure}[!htb]
\centering \scalebox{0.9}[0.8]{\includegraphics[bb=2.9cm 0cm 18.1cm
16.1cm, clip=false]{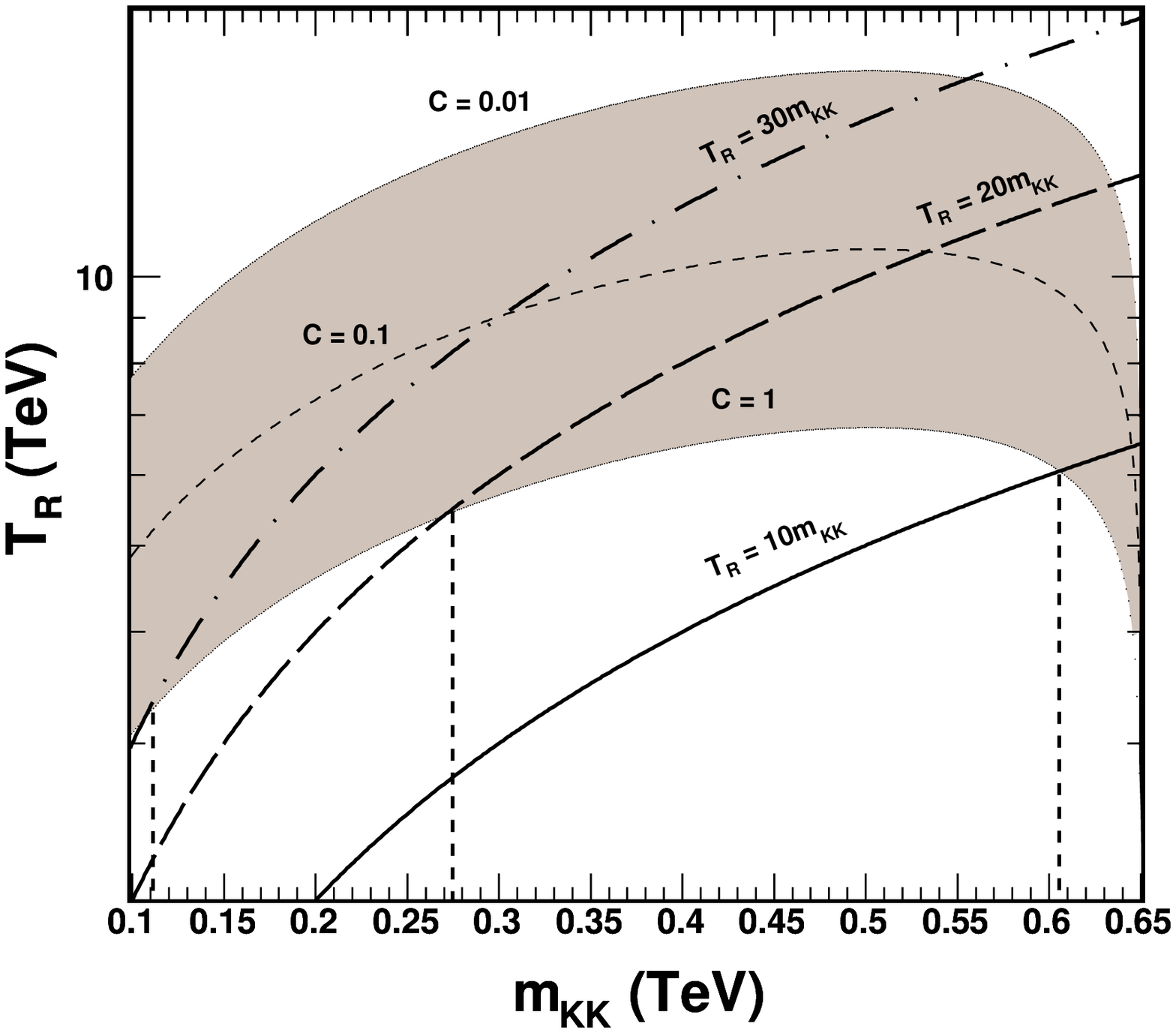}} \caption{\footnotesize{Values of
the reheating temperature $T_R$ consistent with the requirement
$\Omega_{B^1}=\Omega_{\rm DM} =0.23$, for $C=1$, $0.1$, $0.01$, and
assuming $D=6$. Also shown in the Figure are curves of constant
ratio of the reheating temperature to the lightest KK mode, $T_R=10
\; m_{\rm KK}$, $20 \; m_{\rm KK}$ and $30 \; m_{\rm KK}$.}
}\label{fig:TrD6}
\end{figure}

\begin{eqnarray}\label{eq:YB1}
Y_{B^1}&=&dY_\infty+Y_G\\
n_{B^1}&=&s_0Y_{B^1}\\
\Omega_{B^1}&=&\frac{m_{B^1}n_{B^1}}{\rho_c}
\end{eqnarray}
where $\rho_c=5.3\times10^{-9}\mbox{ TeV/}\mbox{cm}^3$, and
$Y_\infty$ is the abundance of the LKP without the inclusion of the
gravitons, Eq.~(\ref{eq:Y_infty}), and $Y_G$ is the abundance of the
gravitons, Eq.~(\ref{eq:YG}). We have not included either
coannihilation or second KK resonance effects in the calculation of
$Y_\infty$.

Requiring that $\Omega_{B^1} \simeq 0.23$, we get :
\begin{equation}\label{eq:Tr_const}
T_R \simeq m_{\rm KK}\left[\frac{\Omega_{B^1} \rho_c-s_0m_{\rm
KK}dY_\infty}{\alpha(d)Cs_0m_{\rm KK}^2}\right]^{\frac{2}{4+3d}}
\end{equation}

In principle, the above expression shows that the relic density
constraint can be satisfied for any value of the lightest KK mode
mass, $m_{\rm KK}$, provided the value of $T_R$ is adjusted
according to Eq.~(\ref{eq:Tr_const}). In practice,
Fig.~\ref{fig:TrD5} shows that in order to get a consistent dark
matter density for low values of $m_{\rm KK}$, one would need large
values of the ratio of the reheating temperature to $m_{\rm KK}$,
but as is discussed in the next section, these are disfavored due to
strong coupling constraints. Similar results are presented in
Fig.~\ref{fig:TrD6} for the case of $D=6$.

The mass of the LKP that would be obtained without the graviton
tower contribution, $m_{\rm WG}$, is shown in the figure as the
point where the bound on $T_R$ becomes very strong, $m_{\rm KK}
\sim0.92$ TeV for $D=5$ and $m_{\rm KK} \sim0.65$ TeV for $D=6$,
since in this case the reheating temperature must become smaller
than the lightest KK mass so that the relic density is not effected
in any significant way by the gravitons.

It is important to stress that for $D=5$, values of the KK masses
smaller than about 500 GeV, are also restricted in the minimal UED
case by precision electroweak constraints~\cite{Appelquist:2000nn,
Appelquist:2002wb,Flacke:2005hb,Gogoladze:2006br}.
Somewhat lower masses may be obtained for large
values of the Higgs mass. However even then, within this minimal
framework, values of $m_{\rm KK}$ smaller than 350~GeV are strongly
disfavored. From
Fig.~\ref{fig:TrD5} we see that for $D=5$, $C\sim 1$ and values of
the reheating temperature $T_R \leq 40\; m_{\rm KK}$ (consistent
with the strong coupling bounds discussed in the next section), a
consistent relic density may be obtained for any value of $m_{\rm
KK}$ larger than 580 GeV. This range of values of the LKP mass are
also consistent with precision electroweak constraints. Even for $C
\sim 0.1$ and for the same range of values for the reheating
temperature there could be modifications of $m_{\rm KK}$ up to ten
percent of the value obtained without the inclusion of gravitons.
Larger (smaller) modifications are possible for larger (smaller)
values of $T_R$.

As shown in Fig.~\ref{fig:TrD6}, for $D=6$ one could obtain larger
effects for smaller values of the ratio of $T_R/m_{\rm KK}$.
However, as will be shown in the next section, the bounds obtained
from the requirement of perturbative consistency of the theory
become much more stringent in this case. Therefore, large departures
from the $m_{\rm KK}$ values obtained in the absence of gravitons
seem to be disfavored for $D=6$ within this minimal framework.

\subsection{Constraints on the Reheating Temperature}

Large values of the reheating temperature compared to $m_{\rm KK}$
immediately raise the question of the ultraviolet cut-off for our
effective 4D theory. In Ref.~\cite{Appelquist:2000nn}, the authors
assumed the limit on the KK masses to be the order of $\sim
40\;m_{\rm KK}$, based on the quantum corrections to the strong
gauge coupling. More stringent bounds have been evaluated in
\cite{Chivukula:2003kq, Chivukula:2004qh, Masip:2000xy}. Here we
show that, if one computes the running of the zero mode hypercharge
gauge coupling, and require it to remain weak in the ultraviolet
regime, one obtains a bound similar to the one obtained in
Ref.~\cite{Appelquist:2000nn}.

Let us then analyze the running of the zero mode gauge couplings for
one extra dimension:
\begin{eqnarray}
g_j^{-2}(\mu)&=&g_j^{-2}(m_z) - \frac{b_j}{8\pi^2}\ln\frac{\mu}{m_z}
-\frac{\tilde{b}_j}{8\pi^2} (N_{KK}\ln[N_{KK}] - \ln[N_{KK}!])\label{eq:KKcoupling}\\
&&\nonumber\\
\mbox{where   }&&\nonumber\\
\tilde{b}_j&=&\left(\frac{81}{10}, \frac{(-44+48+1+2d)}{6}, \frac{(-22+16+d)}{2}\right), \mbox{   each KK level}\nonumber\\
b_j&=&\left(\frac{41}{10},-\frac{19}{6},-7\right)\mbox{
SM}\label{eq:beta}
\end{eqnarray}

For $D=5$, the most dangerous one for the KK scenario was the U(1)
case and we found that $g_1$ develops a Landau pole at scales
$\mu>46 \; m_{\rm KK}$~\cite{Peskin:1995ev,
Ramond:1999vh,Jones:1981we}. Our conclusion is then that above
energy scales of about $40\;m_{\rm KK}$ our effective theory breaks
down. Similarly, for $D=6$, a Landau pole would develop at scales
larger than about $10 \;m_{\rm KK}$. However, these bounds may be
avoided by assuming that this theory is just the low energy
manifestation of a theory with more degrees of freedom and based on
a higher, asymptotically free, gauge group. This would not only
prevent $\alpha_1$ from developing a Landau pole, but additionally
it would increase the value of $C$. We have conservatively kept
values up to $T_R\sim100\;m_{\rm KK}$ for $D=5$ and
$T_R\sim30\;m_{\rm KK}$ for $D=6$.

The impact of these bounds on the possible values of the lightest KK
mass are displayed in Figs.~\ref{fig:TrD5} and~\ref{fig:TrD6}. As we
mentioned before, we see that if we restrict ourselves to the bounds
implied by perturbative consistency, the mass of the LKP can be as
low as $m_{\rm KK}\sim0.6$ TeV for both $D=5$ and $D=6$, but lower
values, as low as those implied by consistency with precision
electroweak measurements, would be only obtained by the possible
relaxation of these bounds by new physics. On the other hand, if the
most stringent bounds obtained in \cite{Chivukula:2003kq,
Chivukula:2004qh, Masip:2000xy} were imposed, very small
modifications to the mass of the LKP would be obtained.

\subsection{Additional Contributions to the Annihilation
Cross Section}

\begin{figure}[!tb]
\centering \scalebox{0.9}[0.8]{\includegraphics[bb=2.9cm 0cm 18.1cm
16.1cm, clip=false]{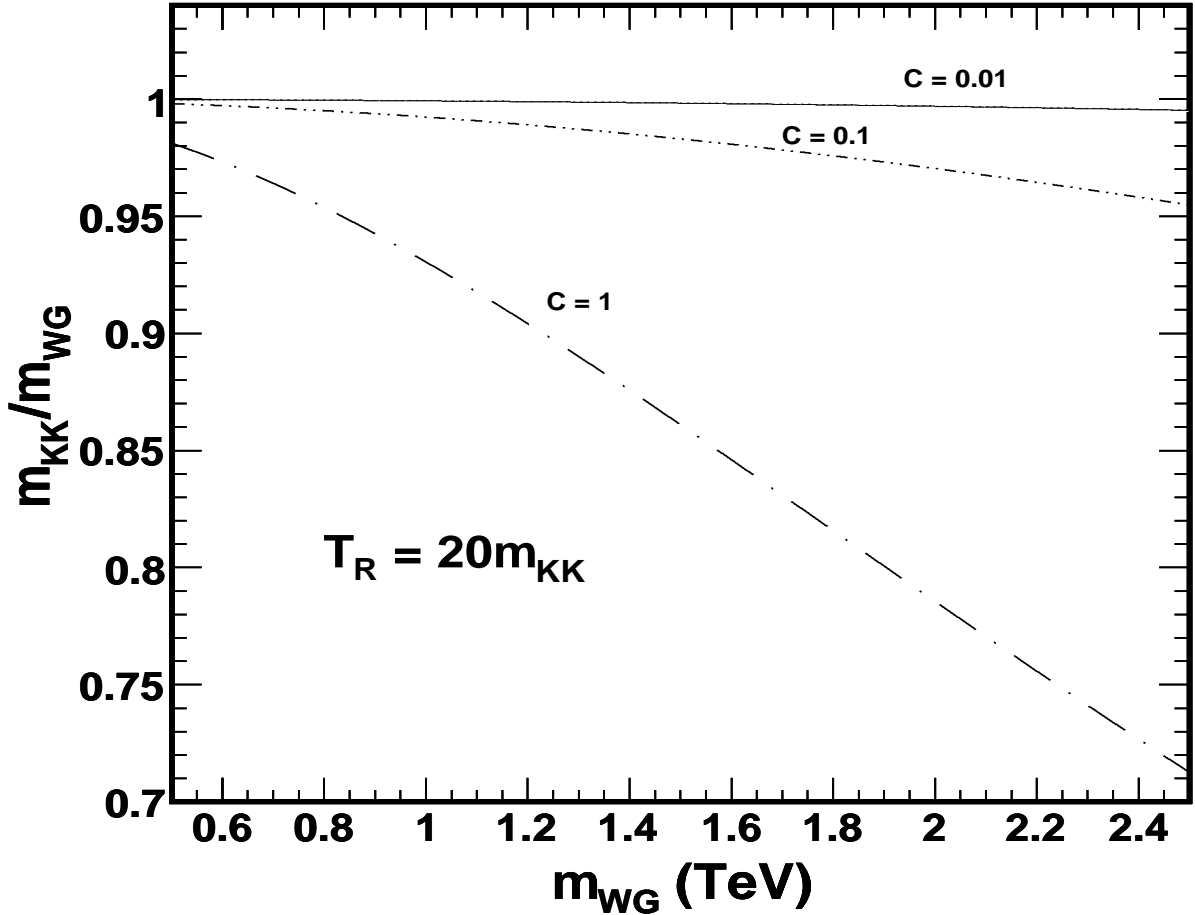}} \caption{\footnotesize{Values of the
ratio of the LKP mass consistent with $\Omega_{\rm DM} = 0.23$ to
the one obtained in the absence of gravitons, $m_{\rm WG}$, for
$T_R=20 \; m_{\rm KK}$, $D=5$ and for $C=0.01$, $0.1$ and $1$}.}
\label{fig:mwg20}
\end{figure}

\begin{figure}[!tb]
\centering \scalebox{0.9}[0.8]{\includegraphics[bb=2.9cm 0cm 18.1cm
16.1cm, clip=false]{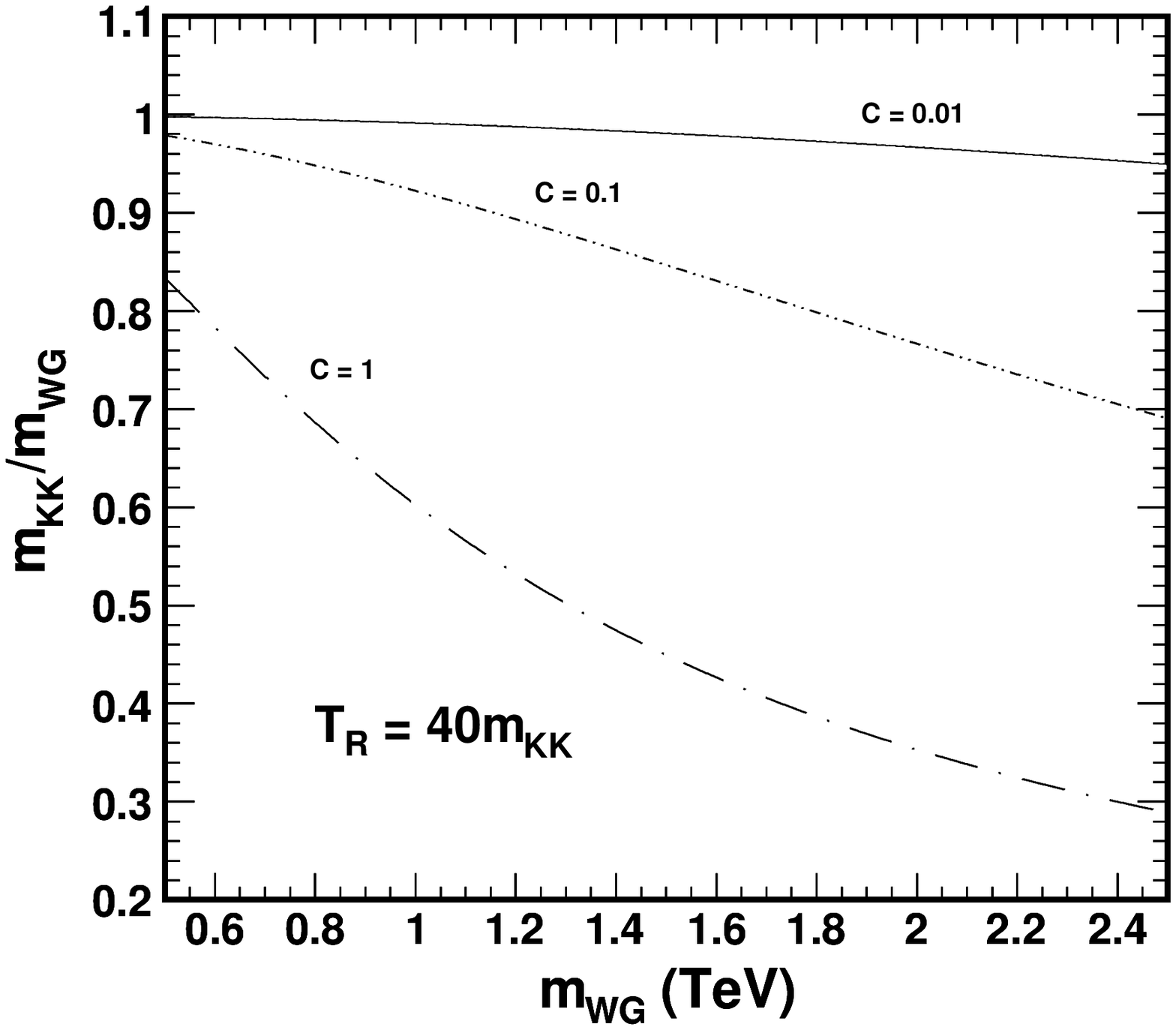}} \caption{
\footnotesize{
Same as Fig.~\ref{fig:mwg20},
but for $T_R=40 \; m_{\rm KK}$}.}
\label{fig:mwg40}
\end{figure}

\begin{figure}[!tb]
\centering \scalebox{0.9}[0.8]{\includegraphics[bb=2.9cm 0cm 18.1cm
16.1cm, clip=false]{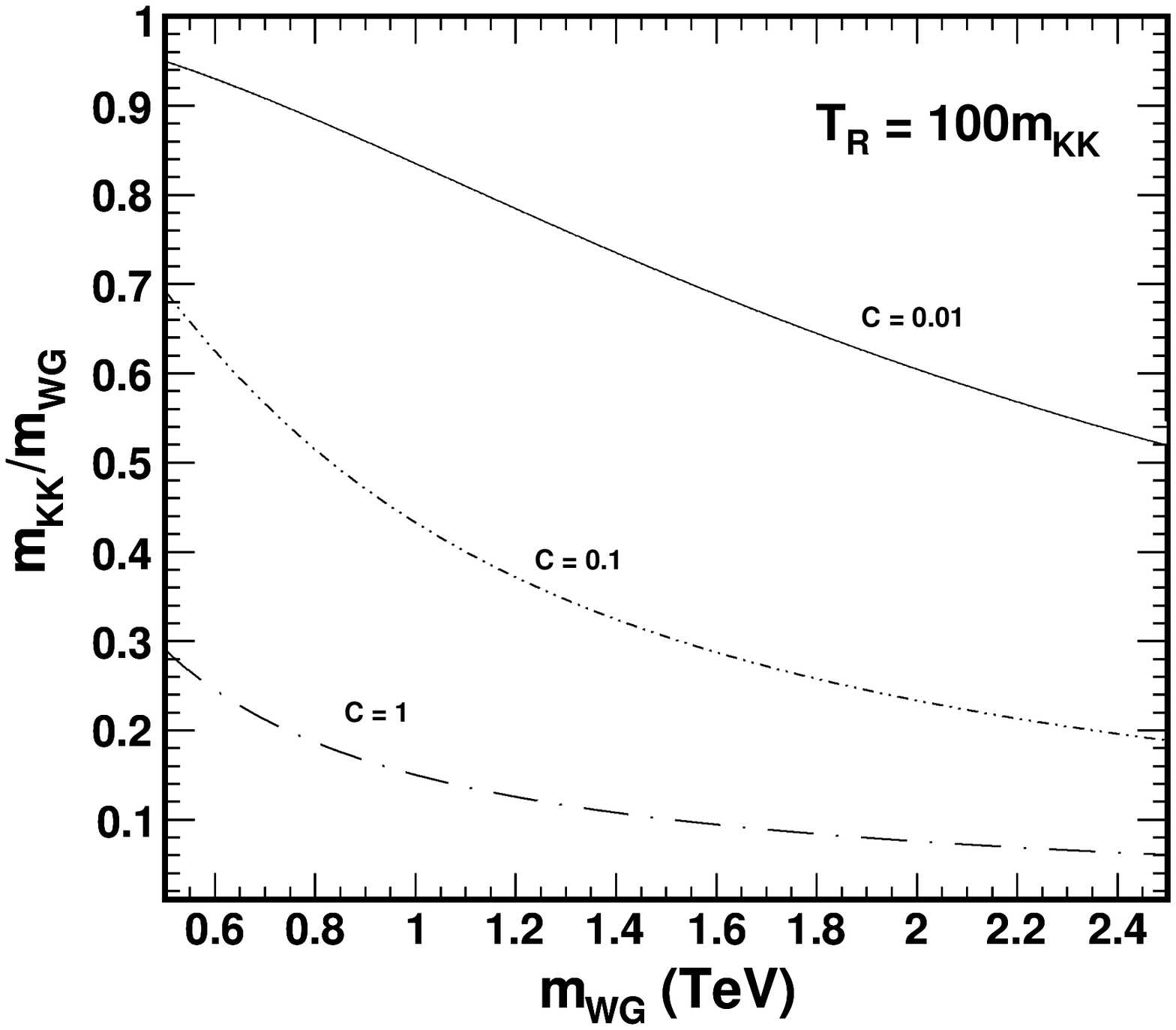}} \caption{\footnotesize{
Same as Fig.~\ref{fig:mwg20}, but for
$T_R=100 \; m_{\rm KK}$ }.}
\label{fig:mwg100}
\end{figure}

Until now, we have only considered the effects induced by the
annihilation of $B^1$s on the total relic density. However, several
effects can change the density of the $B^1$s. For example, if we
assume that the mass difference between the LKP and the NLKP is less
than about 10\%, the NLKP is still thermally accessible to the LKP,
and the effective cross-section, including coannihilations must be
used in Eq.~(\ref{eq:Boltz}). Also, possible resonant effects
induced by the second KK-level modes must be included~\cite{
Griest:1990kh,Kakizaki:2005en,Kakizaki:2005uy,Kong:2005hn,Kakizaki:2006dz}.
Note that even though including these effects changes the mass of
the LKP consistent with the dark matter density without the
inclusion of the gravitons, $m_{\rm WG}$, and will certainly change
any numerical results we obtain, the qualitative picture we have
presented here will remain unchanged.

The corrections to the cross-section can be quantified by
considering $a'$ and $b'$ in Eq.~(\ref{eq:Y_infty}) to encode all
the information about the processes that affect $Y_\infty$. Due to
the weak, logarithmic dependence of $x_F$ on $m_{\rm KK}$ and the
interaction cross section, we can approximately parameterize
$Y_\infty$  as being proportional to $m_{\rm KK}$,
\begin{equation}
\label{eq:y_par} Y_\infty=y\left(\frac{m_{\rm
KK}}{\mbox{TeV}}\right).
\end{equation}
Then $y$, such that $m_{\rm WG}$ (including coannihilation, second
KK resonances etc., but without the graviton tower) is consistent
with the experimentally observed dark matter density, $\Omega_{DM}$,
is given by:
\begin{equation}
\label{eq:y} y=\frac{\rho_c\Omega_{DM}}{s_0 d m^2_{\rm WG}}.
\end{equation}

We can now replace $Y_\infty$ using Eq.~(\ref{eq:y_par}) and
(\ref{eq:y}) in Eq.~(\ref{eq:YB1}). Using this, we get the change in
$m_{\rm WG}$ necessary to reproduce the observed dark matter density
for a given reheating temperature $T_R$:

\begin{equation}
\label{eq:m_wg} \frac{m_{\rm KK}}{m_{\rm
WG}}=\left(\frac{\Omega_{DM}\rho_c}{\Omega_{DM}\rho_c+\alpha(d)Cs_0m^2_{\rm
WG}\left[\frac{T_R}{m_{\rm
KK}}\right]^{\frac{4+3d}{2}}}\right)^{1/2}
\end{equation}
This is plotted in Fig.~\ref{fig:mwg20}, ~\ref{fig:mwg40} and
\ref{fig:mwg100} for $T_R=20\;m_{\rm KK},\;40\;m_{\rm KK}$ and
$100\;m_{\rm KK}$ and a range of $m_{\rm WG}$ predicted in
Refs.~\cite{
Griest:1990kh,Kakizaki:2005en,Kakizaki:2005uy,Kong:2005hn,Kakizaki:2006dz}.
We see that even for $T_R=20\;m_{\rm KK}$ with $C=1$, the mass
consistent with the observed relic density undergoes a relevant
modification due to the inclusion of the gravitons. This effect
becomes increasingly important with increasing $m_{\rm WG}$.
Therefore, depending on the reheating temperature, any precise
calculation of $m_{\rm KK}$ must include the contribution from the
graviton tower to be accurate.

It is interesting to observe that for $T_R \simeq 100 \; m_{\rm KK}$
and $C \simeq 1$, the necessary values of $m_{\rm KK}$ are below the
bounds imposed by precision electroweak constraints. Conversely,
this shows that for this value of $C$ and for values of the KK
masses consistent with precision electroweak constraints, such large
values of the reheating temperature will lead to an excess of dark
matter density and are therefore disfavored. On the contrary, for
values of $T_R \simlt 40 m_{\rm KK}$, the obtained values of $m_{\rm
KK}$ are in good agreement with the scale set by precision
electroweak data. This shows an interesting correlation between the
dark matter density bounds and those coming from requiring the
perturbative consistency of the theory up to scales of the order of
$T_R$.

\subsection{Constraints on the $G^1$--$B^1$ Mass Difference for $\delta'=0$}

In the discussion above, we have ignored the bounds on the energy
released in the graviton decays. As we have stated in section 4, the
amount of energy released will depend on the type of mass correction
the graviton and $B^1$ receive (see Eq.~(\ref{eq:Erel}) and
Eq.~(\ref{eq:xi_bound})). For the case of $\delta'=0$, the energy
released per background photon is given by

\begin{figure}[!tb]
\centering \scalebox{0.9}[0.8]{\includegraphics[bb=2.9cm 0cm 18.1cm
16.1cm, clip=false]{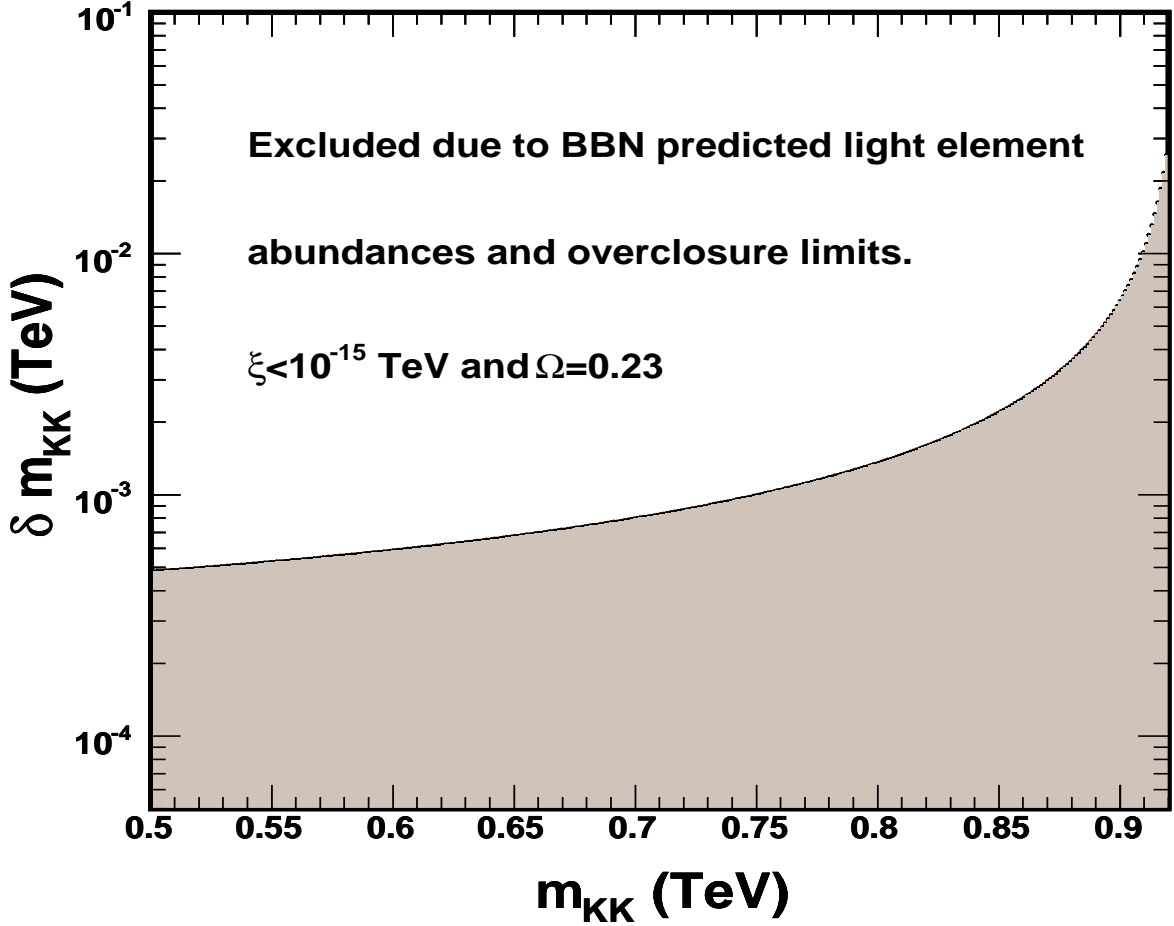}} \caption{\footnotesize{
Values of the allowed mass difference between $G^1$ and the LKP,
$\delta m_{\rm KK}$, for $\delta' = 0$, assuming
$\Omega_{\rm DM}=0.23$ and $\xi<10^{-15}$ TeV for $D=5$. We have
ignored possible coannihilation and second KK level effects.}}
\label{fig:delta1}
\end{figure}

\begin{equation}\label{eq:ERel1}
\xi=\frac{B_{\rm EM/Had}}{n_\gamma}\; \delta m_{\rm KK} \; s_0 \;
Y_G <\xi_B=10^{-15}\mbox{ TeV}.
\end{equation}
Assuming, as stated before, that there are no corrections to the KK
graviton masses $m_{G^n}$, then $m_{G^n}=nm_{G^1}$. In addition we
can identify $m_{\rm KK}$ with $m_{B^1}$. The corrections to
$m_{B^1}=m_{G^1}(1-\delta)$ are then equivalent to corrections to
$m_{G^n}=nm_{B^1}(1+\delta)$ for small $\delta$.

If we now assume that the reheating temperature is such that one
reproduces the correct dark matter relic density,
$\Omega_{B^1}=0.23$, as obtained in the previous section, then we
get a constraint on the mass difference between $G^1$ and $B^1$:

\begin{equation}\label{eq:delta_const}
\delta m_{\rm KK} <\frac{n_\gamma \xi_B}{B_{\rm
EM/Had}(\rho_c\Omega_{B^1}-m_{\rm KK}s_0dY_\infty)}.
\end{equation}
This is plotted in Fig.~\ref{fig:delta1} for $D=5$. This constraint
is independent of $C$. The mass difference between $B^1$ and $G^1$
is very tightly constrained to be less than a GeV for most of the
mass range considered. As $m_{\rm KK}$ approaches the value of the
mass consistent with a proper dark matter relic density in the
absence of gravitons, the bounds become very weak. This is due to
the fact that in this case, as mentioned before, $T_R$ may be of the
order of (or smaller than) $m_{\rm KK}$, and hence very few
gravitons would be produced, implying that the mass difference can
be large without inducing any dramatic effects.

For most of the parameter space, we have derived here a very
stringent constraint on the mass difference. Therefore, we need to
consider whether such small mass differences would require too much
fine tuning to be considered natural. As is discussed in
Section~6.7, we find that in fact the mass differences obtained here
are the order of magnitude of the one loop corrections induced to
the $B^1$ mass.

\subsection{$\delta=0$}

 \begin{figure}[t]
\centering \scalebox{0.9}[0.8]{\includegraphics[bb=2.9cm 0cm 18.1cm
16.1cm, clip=false]{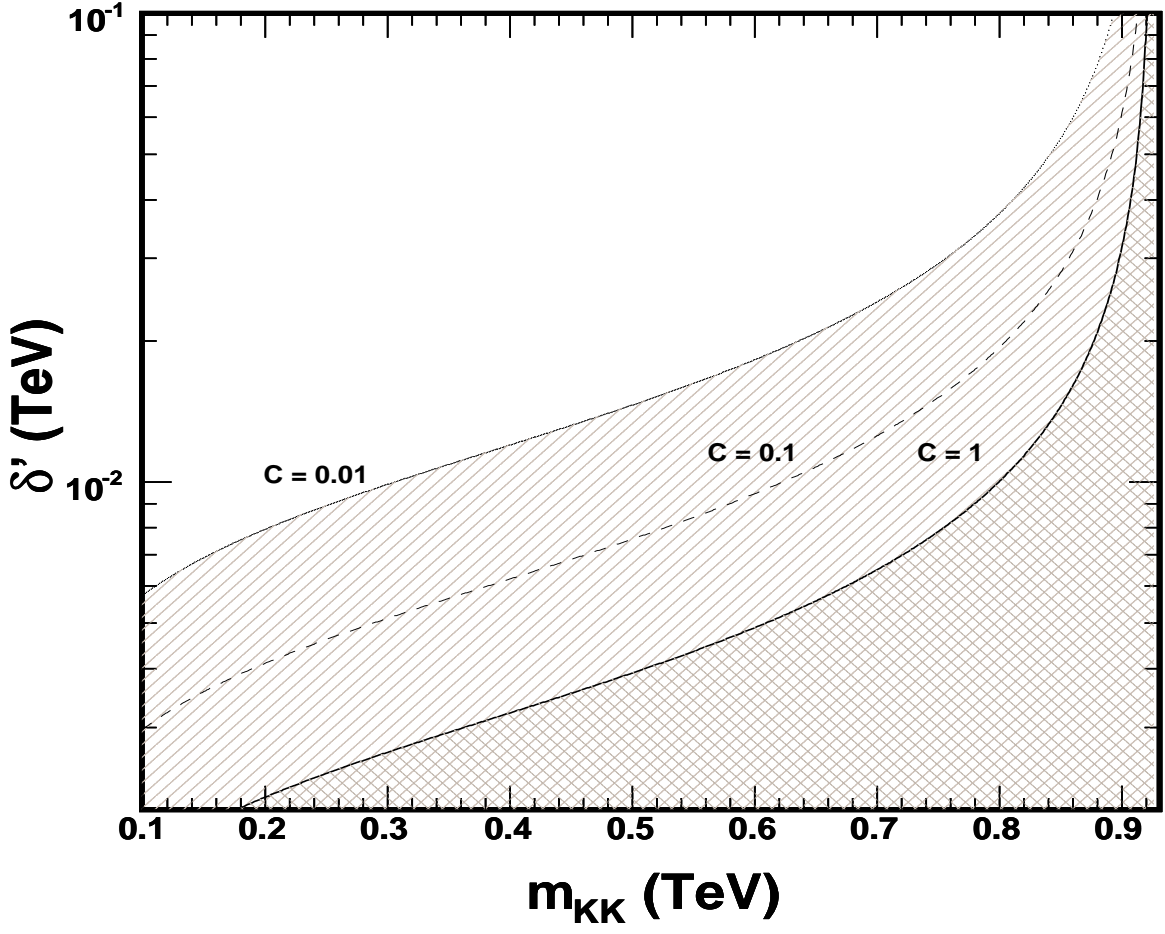}} \caption{\footnotesize{
Values of $\delta'$ if
$\Omega_{\rm DM} =0.23$ and $\xi<10^{-15}$ TeV for $D=5$, $C=1$, $0.1$ and
$0.01$. We have ignored possible coannihilation and second KK level effects.
}} \label{fig:delta2}
\end{figure}

Here we depart from the previous assumption and assume that all
graviton KK modes receive a constant, positive mass correction with
respect to $n m_{\rm KK}$. The energy released for this kind of mass
correction tends to be much smaller than in the case of $\delta \neq
0$ analyzed above (see Eq.~(\ref{eq:ERel1})), and is given by:

\begin{eqnarray}
\xi&=&\frac{B_{\rm EM/Had}}{n_\gamma}\delta' s_0Y_G'<\xi_B=10^{-15}\mbox{ TeV}\label{eq:ERel2}\\
Y_G'&=&\int Y_{G^n}d^dn\nonumber\\
    &=&\frac{45\sqrt{5}}{2\pi^8}\zeta^2(3)\alpha_3\frac{m_{\rm KK}}{M_4}C\sqrt{g_*}\sqrt{\frac{V_dA_d^2}{2^{3d}}}\frac{3+d}{d(2+3d)}\left[\frac{T_R}{m_{\rm KK}}\right]^{1+\frac{3d}{2}}\nonumber\\
    &=&\beta(d)\;C\;\frac{m_{\rm KK}}{\mbox{TeV}}\;\left[\frac{T_R}{m_{\rm KK}}\right]^{1+\frac{3d}{2}}\label{eq:YG'}.
\end{eqnarray}

For this case, since the dependance of $Y_G'$ on $T_R$ is different
from that of $\Omega$, we get a more complicated expression for
$\delta'$, depending explicitly on the number of extra dimensions
and $C$:

\begin{equation}
\delta' <\frac{n_\gamma \xi_B}{B_{\rm EM/Had}\beta(d)Cs_0m_{\rm
KK}}\left[\frac{\Omega_{B^1}\rho_c-s_0m_{\rm
KK}dY_\infty}{\alpha(d)s_0Cm_{\rm
KK}^2}\right]^{-\frac{2+3d}{4+3d}}\label{eq:delta'}
\end{equation}

This is plotted in Fig.~\ref{fig:delta2}. As anticipated, the bounds
on $\delta'$ are much weaker than on $\delta m_{\rm KK}$. However,
even though the constraint is much weaker now than before, only
small mass differences, smaller than about 10 GeV, are allowed in
most of the parameter space.

\subsection{Diffuse Photon Flux Constraints on the $G^1-B^1$ Mass Difference}

Since the $G^1$ decays late to the $B^1$ we have to consider its
effect on the diffuse photon flux. As long as the $G^1$ decays after
matter domination, it can have an effect on the diffuse photon flux.
Since the observed flux is smooth to a high degree, we can derive a
constraint on the mass difference $\Delta_1$ between $G^1$ and $B^1$
by requiring that Eq.~(\ref{eq:maxFlux}) be less than
Eq.~(\ref{eq:obsFlux}). Using this inequality, we are led to:

 \begin{figure}[t] \centering
\scalebox{0.9}[0.8]{\includegraphics[bb=2.9cm 0cm 18.1cm 16.1cm,
clip=false]{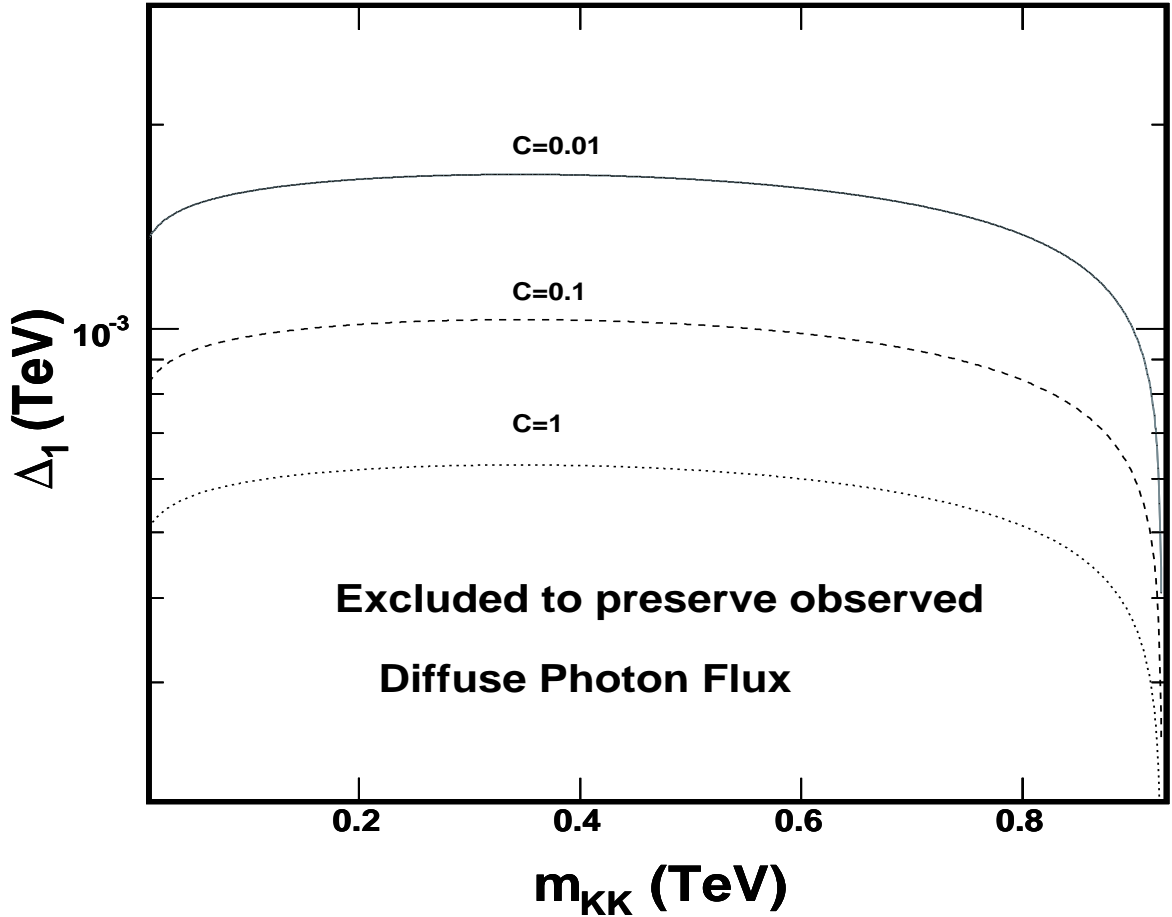}} \caption{\footnotesize{Constraints on the
value of the mass difference between $G^1$ and the LKP, $\Delta_1$,
assuming
$\Omega_{\rm DM} =0.23$ for $D=5$, $C=1$, $0.1$ and $0.01$. We have
ignored possible coannihilation and second KK level effects.}}
\label{fig:delDPF}
\end{figure}

\begin{equation}
\Delta_1\;>\left(2.48\times10^{-3}\left(\left[\frac{T_R}{m_{\rm
KK}}\right]^{3/2}-1\right)\left[\frac{m_{\rm
KK}}{\mbox{TeV}}\right]\right)^{1/2}\;\mbox{GeV},\label{eq:Delta1}
\end{equation}
where $T_R$ is given by Eq.~(\ref{eq:Tr_const}). This is plotted in
Fig.~\ref{fig:delDPF}. We see again that as $m_{\rm KK}$ approaches
the value consistent with the dark matter density without gravitons,
the constraint on $\Delta_1$ becomes very weak. At these masses,
even if the $G^1$ were decaying right now, their density is so small
that the decays are basically invisible and don't affect the photon
spectrum.

\subsection{Comparison of Constraints}

 \begin{figure}[t] \centering
\scalebox{0.9}[0.8]{\includegraphics[bb=2.9cm 0cm 18.1cm 16.1cm,
clip=false]{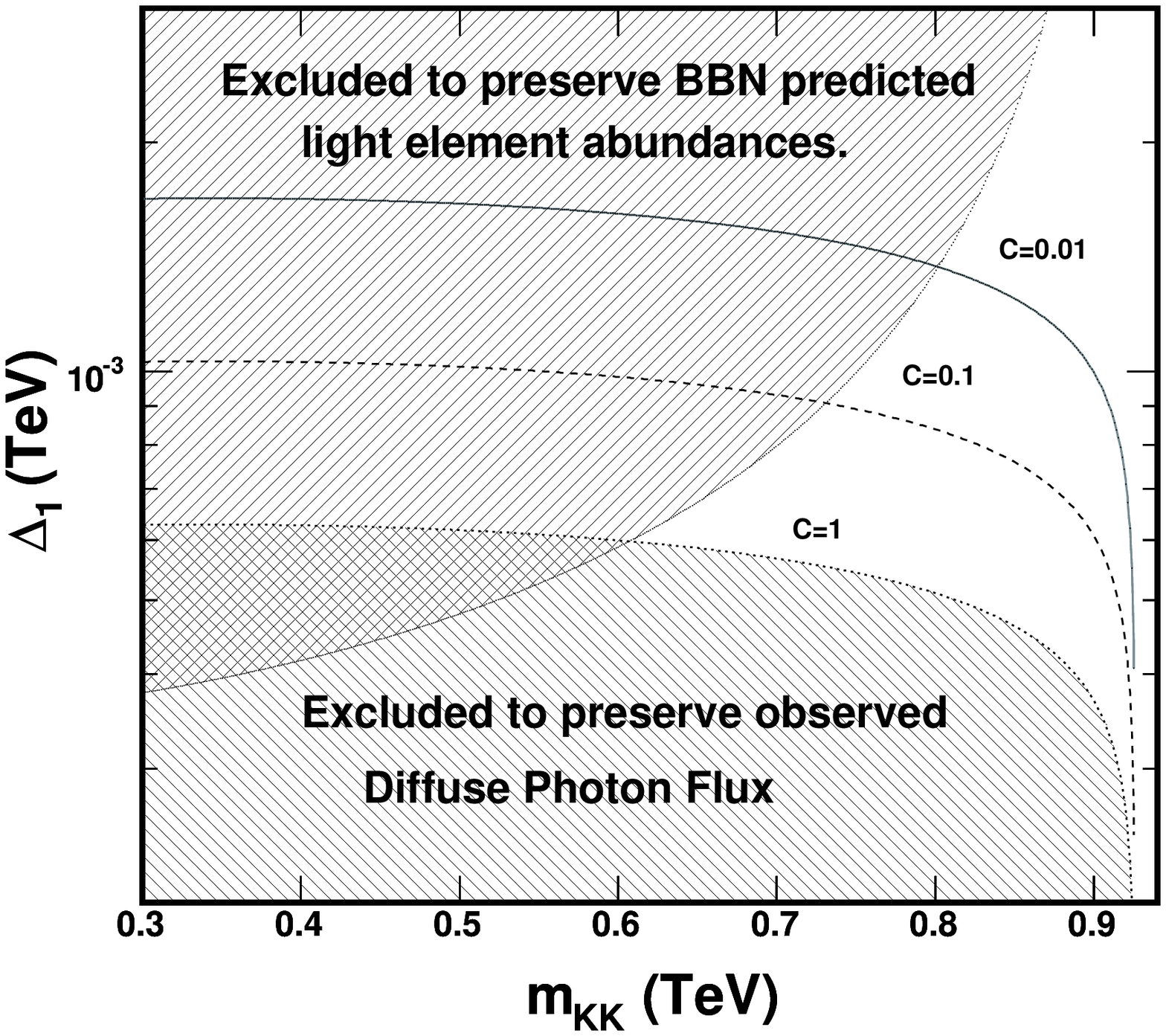}} \caption{\footnotesize{
Constraints on the value of the mass difference between $G^1$ and
the LKP, $\Delta_1$,
assuming $\Omega_{\rm DM}=0.23$ for $D=5$, $C=1$, $0.1$ and $0.01$.
We have ignored possible coannihilation and second KK level effects.}}
\label{fig:delsuper}
\end{figure}

\begin{figure}[t] \centering
\scalebox{0.9}[0.8]{\includegraphics[bb=2.9cm 0cm 18.1cm 16.1cm,
clip=false]{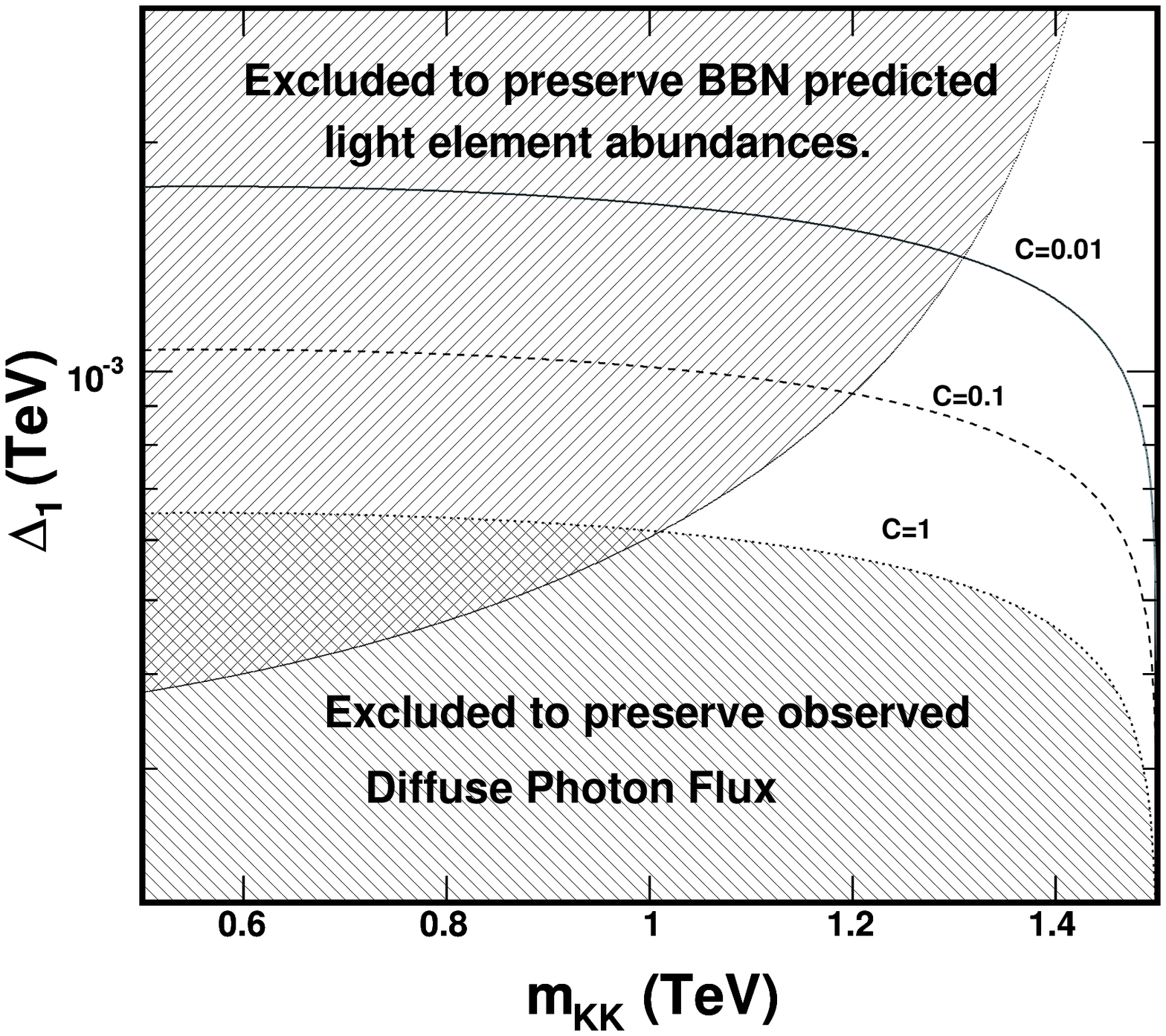}} \caption{\footnotesize{
Same as Fig.~\ref{fig:delsuper}, but for a value of
$m_{\rm WG}=1.5$ TeV.}} \label{fig:delsuper15}
\end{figure}

\begin{figure}[t] \centering
\scalebox{0.9}[0.8]{\includegraphics[bb=2.9cm 0cm 18.1cm 16.1cm,
clip=false]{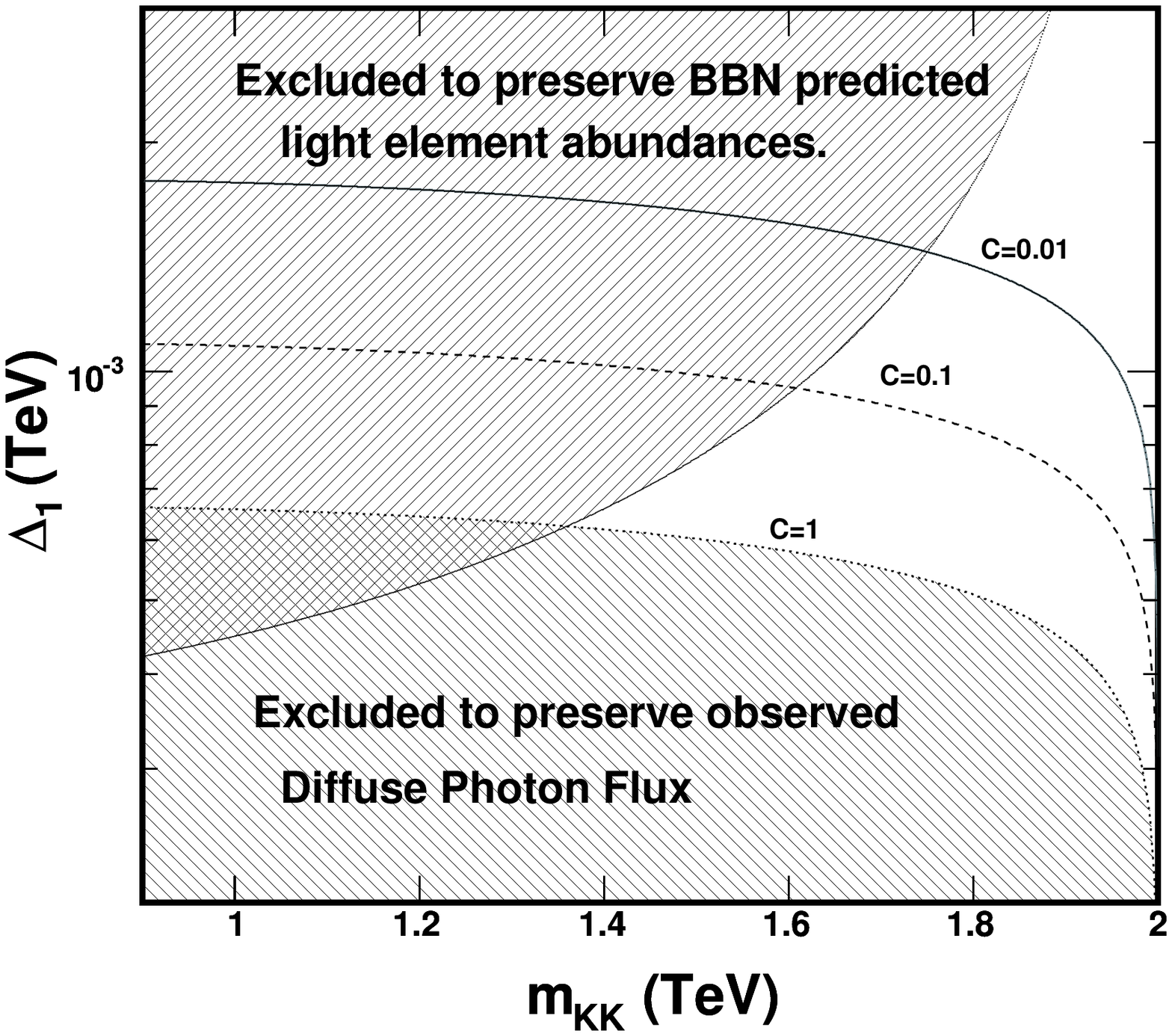}} \caption{\footnotesize{
Same as Fig.~\ref{fig:delsuper}, but for a value of
$m_{\rm WG}=2$ TeV.}} \label{fig:delsuper2}
\end{figure}

Comparing the constraints on the mass difference between $B^1$ and
$G^1$, plotted in Fig.~\ref{fig:delsuper}, we see that there is a
region of parameter space where both conditions are satisfied. The
allowed region increases for larger values of $C$, and for $C = 1$,
extends up to values of order 600~GeV, which is about the value of
the bound placed by precision electroweak tests for a light Higgs
boson. The range of allowed values of $\Delta_1$ shrinks for
decreasing values of $m_{\rm KK}/m_{\rm WG}$. For instance, in the
case represented in Fig.~\ref{fig:delsuper} for $C=1$, values of
$m_{\rm KK} \simeq 0.8$~TeV may be obtained for a range of values of
$\Delta_1\sim\mathcal{O}~(1)$ GeV, while values as low as $m_{\rm
KK}\sim 0.6$~TeV are only allowed for $\Delta_1 \sim 0.6$~GeV.
Interestingly enough the allowed values of $m_{\rm KK}$ agree with
the ones consistent with the observed dark matter relic density for
$T_R \simlt 40 \; m_{\rm KK}$ depicted in Fig.~\ref{fig:TrD5}.

The effects on the allowed values of $\Delta_1$ as $m_{\rm WG}$
changes are illustrated in Figs.~\ref{fig:delsuper15} and
\ref{fig:delsuper2}. As shown in these figures, the qualitative
behavior for the different $m_{\rm WG}$ is the same as the one found
in Fig.~\ref{fig:delsuper}.

\subsection{One Loop Corrections to the KK Masses}

We want to compare the bounds obtained on $\Delta_1$ due to BBN and
the diffuse photon flux found in the previous section, with the
one-loop corrections induced in the $B^1$ mass, assuming that
$m_{G^1}=1/R$. Radiative corrections induced by boundary terms at
the orbifold fixed points affect the spectrum of the standard KK
particles. Since radiative corrections are divergent they may be
regularized by counterterms localized at these boundaries. The
possibility of including localized counterterms also implies that
one could have started with different values of the localized
boundary terms than the ones assumed in the minimal scenario, and
this would lead to modifications of the spectrum different from the
one discussed below~\cite{Carena:2002me}. We include, however, a
discussion of this minimal scenario in order to show that,
interestingly enough, the magnitude of the one-loop induced
corrections is indeed of the order of the one necessary to satisfy
the BBN and diffuse gamma ray bounds.

In order to compute the LKP mass, apart from the one-loop
corrections, one must also take into account the $g_i^2 v^2/4$ correction
to the gauge boson mass matrix due to the vev of the Higgs
field~\cite{Cheng:2002iz}. In the $B^n$, $W^n_3$ basis:
\begin{equation}
\left(\begin{array}{cc}\displaystyle{\frac{n^2}{R^2}+\delta
(m_{B^n}^2)+\frac{1}{4}\;g_1^2v^2}&
\displaystyle{\frac{1}{4}\;g_1g_2v^2}\\
\\
\displaystyle{\frac{1}{4}\;g_1g_2v^2}&\displaystyle{\frac{n^2}{R^2}+\delta (m_{W^n_3}^2)+\frac{1}{4}\;g_2^2v^2}\\
\end{array}\right)
\label{eq:MassMatrix}
\end{equation}

Since the mixing is very small compared to the mass difference
between the $B^n$ and the $W^n_3$ induced at the one-loop level,
the neutral gauge bosons
approximately become pure $B^n$ and $W^n_3$. Therefore, the mass
correction to $B^1$ is very well approximated by:

\begin{equation}\label{eq:Rad_corr}
\delta(m_{B^1}^2)\simeq -
\left[\frac{39}{2}\frac{\alpha_1\zeta(3)}{4\pi^3}+\frac{1}{6}\frac{\alpha_1}{4\pi}\ln\frac{\Lambda^2}{\mu^2}\right]\left(\frac{1}{R}\right)^2+\frac{g_1^2v^2}{4},
\;\;\; \normalsize{{\mbox{D=5}}}
\end{equation}

\begin{figure}[t] \centering
\scalebox{0.9}[0.8]{\includegraphics[bb=2.9cm 0cm 18.1cm 16.1cm,
clip=false]{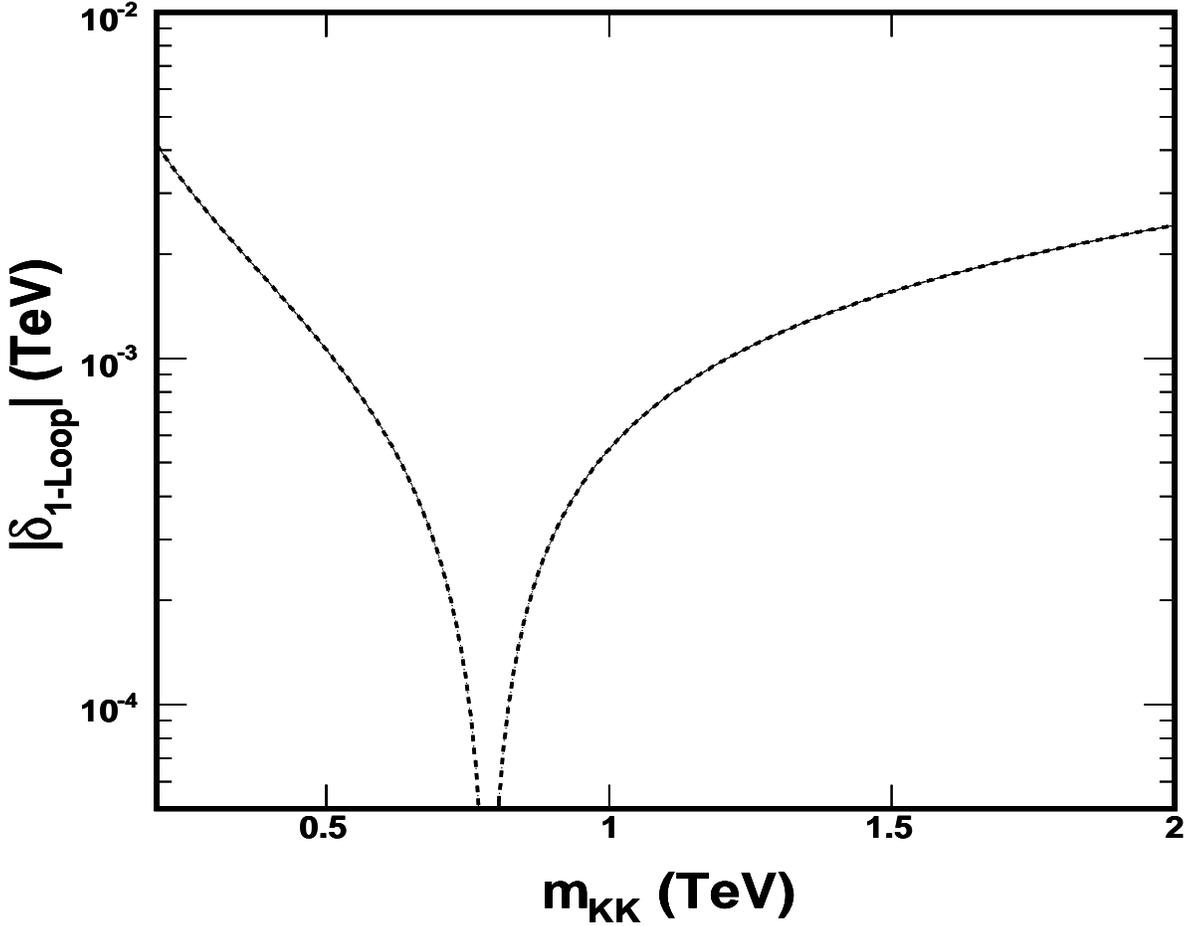}} \caption{\footnotesize{One loop corrections
to the $B^1$ mass as a function of $m_{\rm KK}$.}}\label{fig:1loop}
\end{figure}

This correction to $B^1$ becomes positive for masses below about
$800$ GeV, and the absolute value is plotted in
Fig.~\ref{fig:1loop}. Therefore, as noted in~\cite{Kakizaki:2006dz},
below this mass, assuming only one-loop corrections, the graviton is
the LKP. This case has been extensively studied and the constraints
are derived in Ref.~\cite{Feng:2003xh}. The $B^1$s, with a density
comparable to $\Omega\sim0.23$, would be decaying late to $G^1$ and
based on an analysis similar to the one performed in Section 4, one
concludes that, apart from the finely tuned case in which the mass
difference approximately vanishes, small mass differences between
the $B^1$ and the $G^1$ would lead to a large impact on the diffuse
photon flux. Therefore, unless some evidence is seen in the diffuse
photon flux for new physics for values of
$E_{\gamma}\sim\mathcal{O}~(1)$ MeV (see Fig.~\ref{fig:DPF}), this
minimal scenario would be ruled out for
$m_{\rm KK} < 800$~GeV~\footnote{Please see Ref.~\cite{Matsumoto:2006bf}
for a possible solution to this problem, by the introduction of Dirac
neutrinos.}.

We see that in the allowed region demarcated in
Figs.~\ref{fig:delsuper},~\ref{fig:delsuper15} and
\ref{fig:delsuper2} the values of $\Delta_1$ are, as emphasized
before, of the same order of magnitude as the range predicted by the
one loop corrections shown in Fig.~\ref{fig:1loop}. The precise
quantitative constraints on the possible values of $m_{\rm
KK}/m_{\rm WG}$ depend on $m_{\rm WG}$. While for $m_{\rm WG} \simeq
1$, 2~TeV, only corrections of the order of ten percent would be
allowed, for $m_{\rm WG} \simeq 1.5$~TeV, the one-loop corrections
are consistent with those necessary to satisfy both the BBN and
diffuse photon flux constraints~\footnote{In
Ref.~\cite{Kakizaki:2006dz} it was argued that, within this minimal
framework, values of $m_{\rm WG} > 1.4$~TeV would be disfavored
since a charged Higgs would become the LKP.}.

In evaluating these constraints we have to stress again that we have
used a very conservative estimate for the hadronic branching ratio
of $B_H=1$. The actual energy released into hadrons is approximately
proportional to $B_H$ and would be somewhat lower since the KK
quarks are much heavier then the KK right handed leptons.
Additionally there are large errors in the detection of the diffuse
photon flux. A proper computation of the hadronic branching ratio of
the decay of gravitons, as well as a more accurate spectrum for the
diffuse photon flux, would be necessary in order to determine the
compatibility of this minimal scenario with the energy release
constraints.

\subsection{Determination of the Reheating
Temperature}

\begin{figure}[t] \centering
\scalebox{0.9}[0.8]{\includegraphics[bb=2.9cm 0cm 18.1cm 16.1cm,
clip=false]{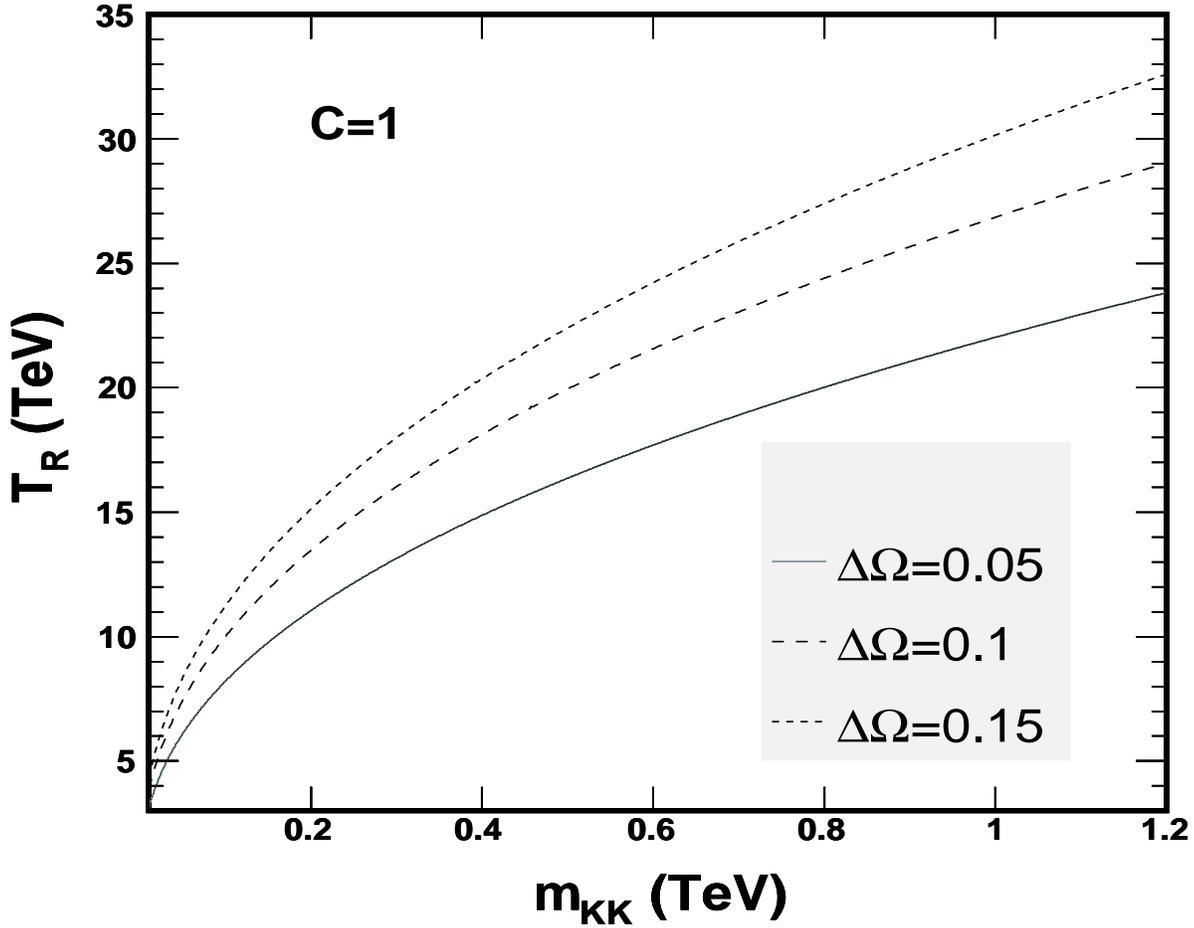}} \caption{\footnotesize{ The reheating
temperature, $T_R$, required to make up the deficit dark matter
density by KK gravitons as dictated by Eq.~(\ref{eq:TrOm}).}
}\label{fig:TrOm}
\end{figure}

Let us finish this section by commenting on the possible
experimental probes of this scenario. Due to their small couplings
to matter, provided $G^1$ is not the LKP, the KK gravitons will not
be produced at laboratory experiments. Therefore, their presence may
only be probed by indirect effects, like their impact on cosmology
as we have discussed in this article. In order to be able to
evaluate these effects, one would have to find conclusive evidence
of UED at, for instance, collider experiments and, in addition,
measure the properties of the relevant first and second KK modes
that contribute to the $B^1$ annihilation cross section. Then, under
the assumption that no other exotic particle contributes to the dark
matter density one could estimate the relic density associated with
the LKP for the specific value of $m_{\rm KK}$ measured. If the
obtained relic density disagrees with the experimentally measured
one, then one would get information about the possible KK graviton
contribution.

Knowledge of the graviton contribution and the KK mass $m_{\rm KK}$
will, in turn, allow us to determine the reheating temperature. By
analogy with Eq.~(\ref{eq:Tr_const}), the reheating temperature for
one extra dimension would be given by,
\begin{equation}
T_R = m_{\rm KK} \left(\frac{\Delta \Omega \; \rho_c}{\alpha(1) C s_0
m_{\rm KK}^2} \right)^{2/7} , \label{eq:TrOm}
\end{equation}
where $\Delta \Omega$ is the difference between the measured value
of $\Omega$ and the theoretically estimated one from the measured
value of $m_{\rm KK}$ and the corresponding annihilation cross
section. This is plotted for $C=1$ in Fig.~\ref{fig:TrOm} for
several different values of $\Delta\Omega$ for reference. Even
including the possibility of exotic particles contributing to the
dark matter density, Eq.~(\ref{eq:TrOm}) gives an upper limit on the
reheating temperature.

\section{Conclusions}

In this article, we analyzed the effects of including the KK
graviton tower on the determination of the relic density associated
with the lightest KK particle in the scenario of Universal Extra
Dimensions. Gravitons may be copiously produced in the early
universe and their subsequent decay into the LKP may lead to a
modification of the observed LKP relic density. Graviton production
is governed by a parameter $C \sim {\cal{O}}(1)$ and by the
reheating temperature, which is bounded from above by the
requirement of maintaining the perturbative consistency of the
theory, $T_R < 40 m_{\rm KK}$ ($T_R < 10 m_{\rm KK}$) for $D =5$
($D=6$). In the case that Universal Extra Dimensions are observed in
laboratory experiments, we show that an upper limit on the reheating
temperature can be deduced from the requirement that only the LKP
contributes to the observed dark matter density.

Throughout this work we have assumed KK parity conservation, leading
to the stability of the LKP particle. We found that including the
graviton effectively lowers the lightest KK particle mass consistent
with the observed dark matter relic density from values of about
0.92 TeV to values as low as $m_{\rm KK}\sim0.58$ TeV for $C=1$,
$D=5$ and $T_R<40\;m_{\rm KK}$. Additionally, including effects
which change the LKP density (coannihilation, second KK-level mode
resonant contributions, etc.), we show that the graviton tower has a
large impact on the predicted mass of the LKP with increasing mass.
It should be stressed here that these results are independent of
which KK particle is the LKP (or the NLKP).

Additionally, there are bounds on the mass difference of the LKP
with the graviton KK modes induced by the requirement that the
energy released in the graviton decay does not lead to a disturbance
of the light element abundance or the diffuse photon flux. Under the
assumption that the graviton spectrum is $m_{G^n} \sim n/R$, we have
obtained a bound on the mass difference of $G^1$ and the LKP ($B^1$)
mass, that is consistent with the minimal one-loop corrections
obtained in Ref.~\cite{Cheng:2002iz} for a large range of values of
$m_{\rm KK}$.
~\\
~\\
~\\
\large{\textbf{Acknowledgements:}} \normalsize We would like to
thank Jonathan Feng, Csaba Balazs and Tim Tait for useful
discussions and comments. Work at ANL is supported in part by the US
DOE, Div.\ of HEP, Contract W-31-109-ENG-38. This work was also
supported in part by the U.S. Department of Energy through Grant No.
DE-FG02-90ER40560.

%\clearpage

\newpage
\appendix
{\Large{{\bf APPENDIX}}}
\renewcommand{\theequation}{A.\arabic{equation}}
% redefine the command that creates the equation no.
\setcounter{equation}{0}  % reset counter

\renewcommand{\thefigure}{A.\arabic{figure}}
% redefine the command that creates the figure no.
\setcounter{figure}{0}  % reset counter

\section{Graviton Production Cross-Section}
\label{appendixA}

We will calculate explicitly the graviton production cross-section
for $q^l(k_1)+\bar{q}^m(k_2)\to g^k(p_1)+G^n(p_2)$. The feynman
rules we will need are \cite{Feng:2003nr,Macesanu:2003jx}:

\begin{figure}
\begin{center}
$\begin{array}{cc} \scalebox{0.33}[0.33]{\includegraphics[bb=2.9cm
0cm 20.1cm 16.1cm, clip=false]{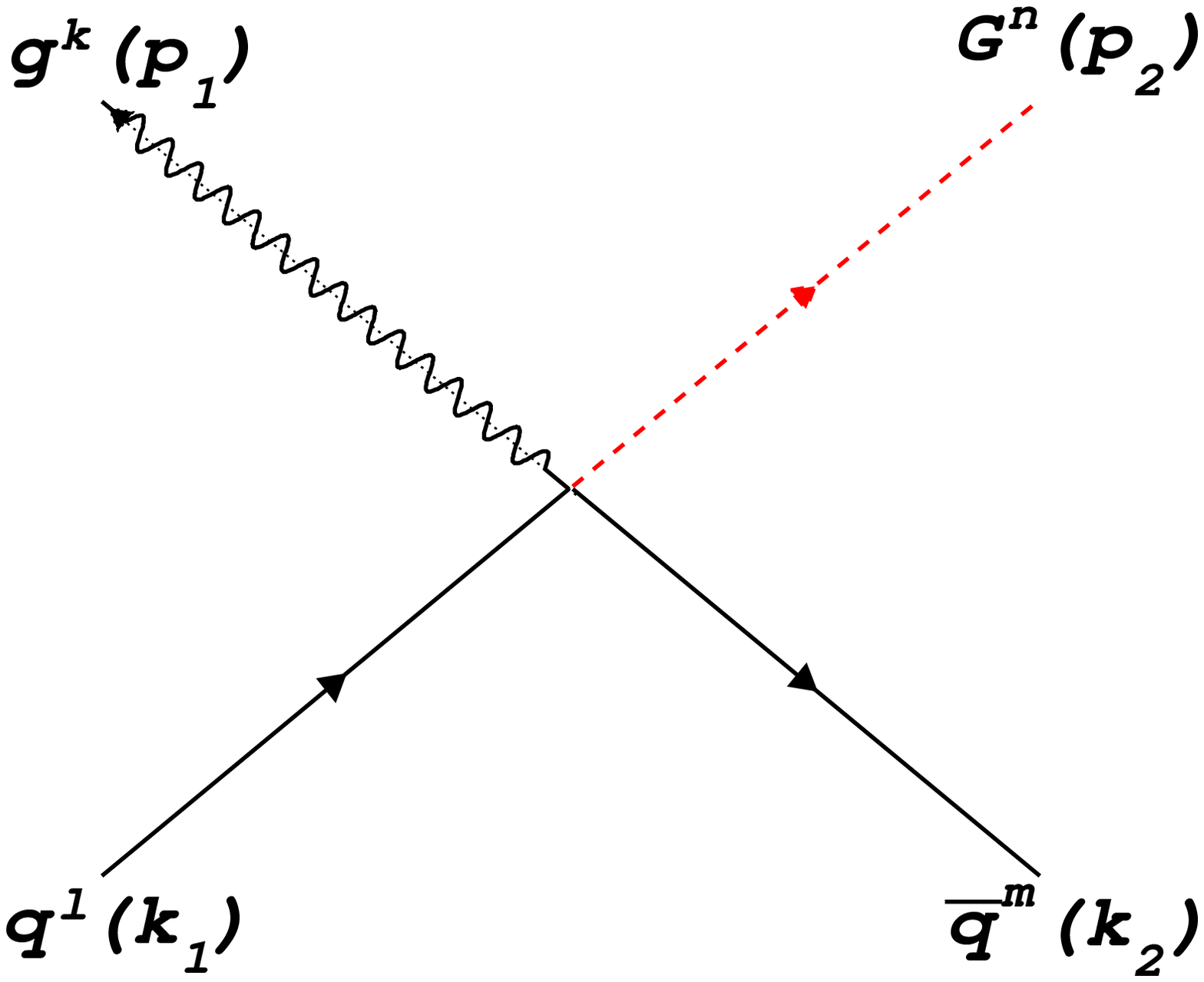}} &
\scalebox{0.3}[0.35]{\includegraphics[bb=0.9cm 2.9cm 18.1cm
16.1cm, clip=false]{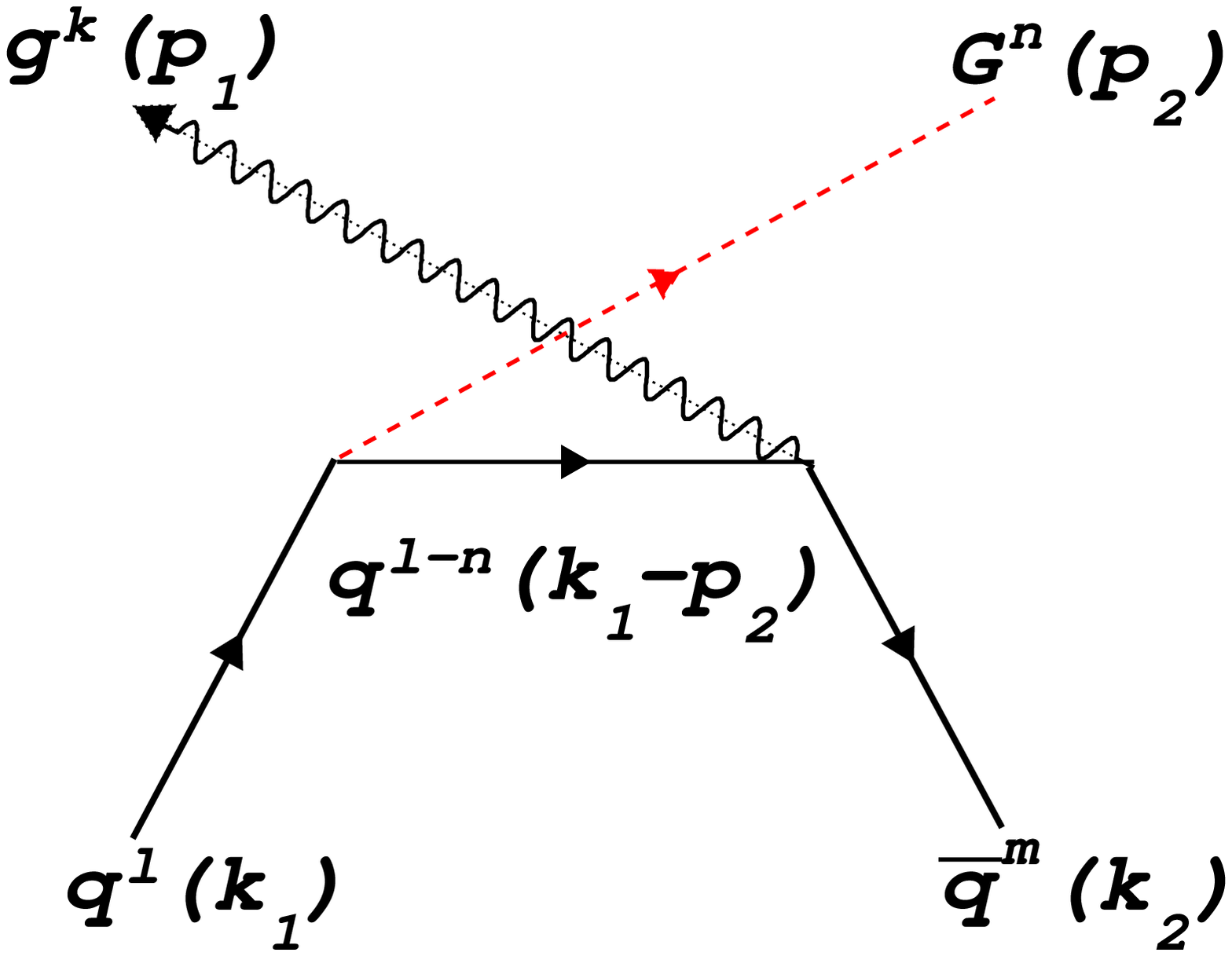}}  \\
\mathcal{M}^1  & \mathcal{M}^2
\end{array}$
\end{center}
\caption{\footnotesize{Feynman Diagrams contributing to
$q^l+\bar{q}^m\to g^k+G^n$. The corresponding amplitudes are given
in (\ref{eq:60}) and (\ref{eq:61}).}}\label{fig:7}
\end{figure}

\begin{figure}
\begin{center}
$\begin{array}{cc} \scalebox{0.33}[0.33]{\includegraphics[bb=2.9cm
0cm 20.1cm 16.1cm, clip=false]{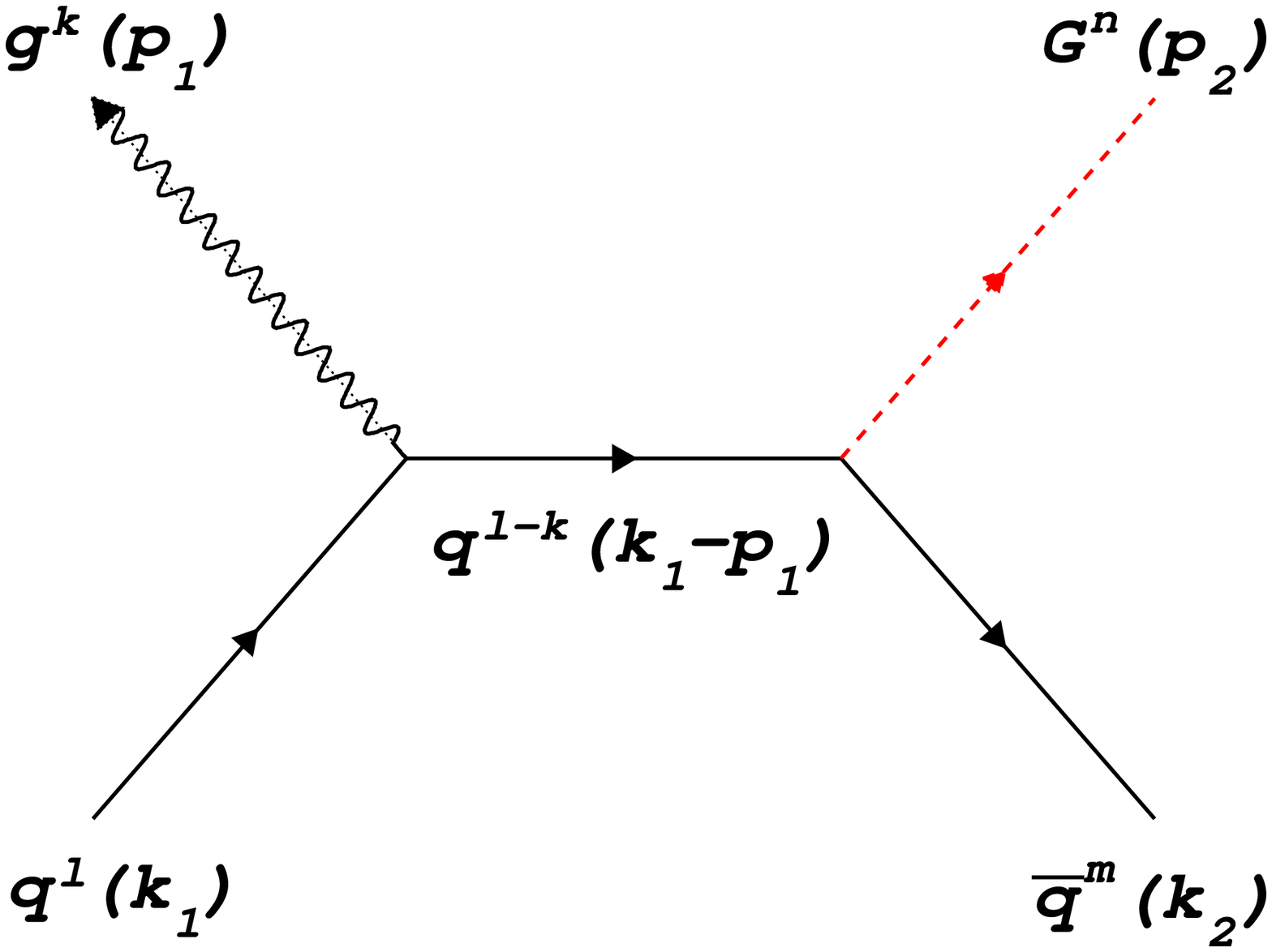}} &
\scalebox{0.3}[0.3]{\includegraphics[bb=0.9cm 0.9cm 18.1cm
16.1cm, clip=false]{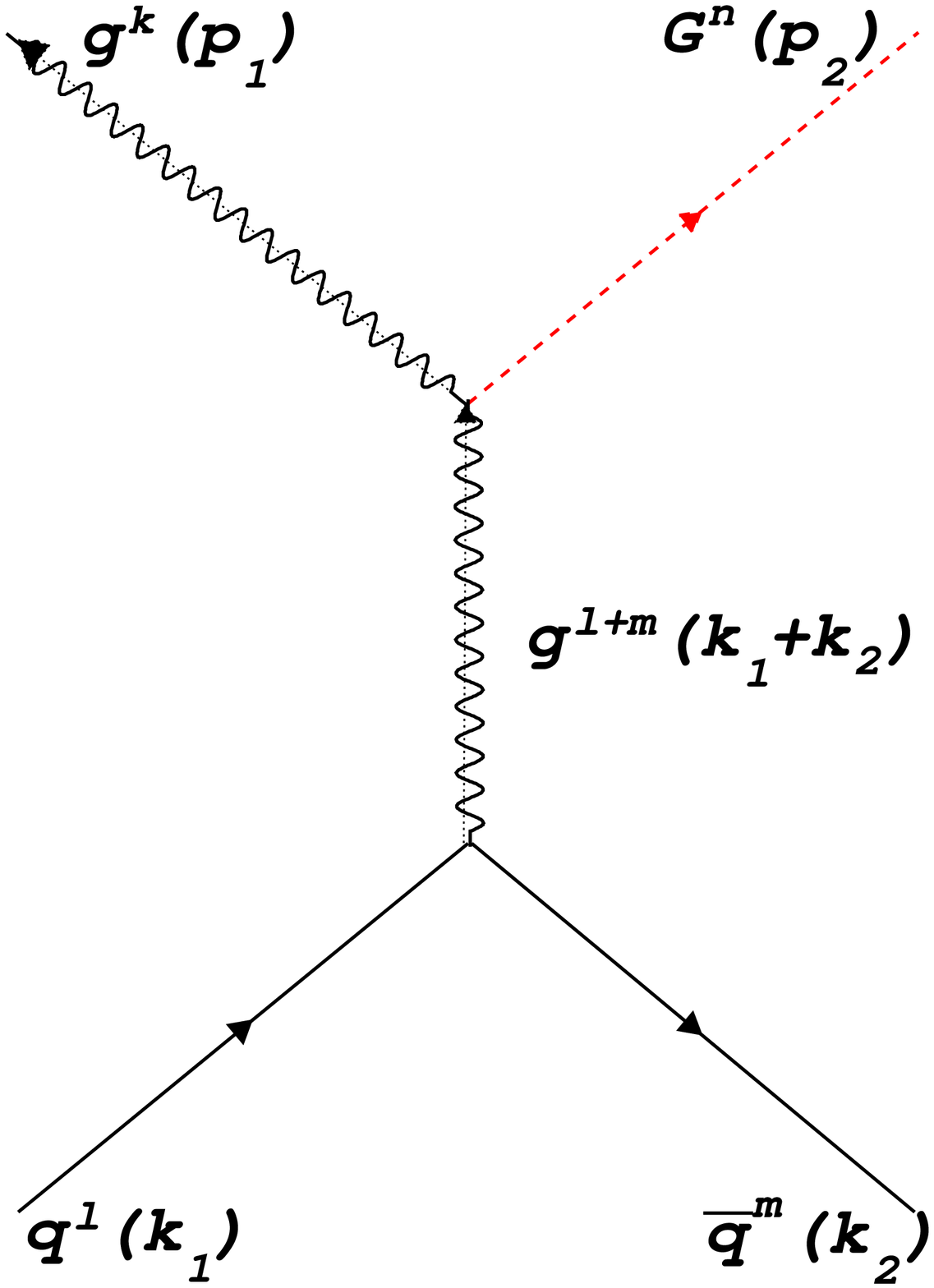}}  \\
&\\
 \mathcal{M}^3 & \mathcal{M}^4
\end{array}$
\end{center}
\caption{\footnotesize{Feynman Diagrams contributing to
$q^l+\bar{q}^m\to g^k+G^n$. The corresponding amplitudes are given
in (\ref{eq:62}) and (\ref{eq:63}).}}\label{fig:8}
\end{figure}

\begin{eqnarray}
q^l(k_1)&\to&
q^m(k_2)+G^n(p_2):\nonumber\\
&&\nonumber\\
Y^{\mu\nu}&=&\frac{i}{4M_4}\left(-\eta^{\mu\nu}(\not{k_1}+\not{k_2}-m_{m\pm l})+\frac{1}{2}\gamma^\mu(k_1+k_2)^\nu+\frac{1}{2}\gamma^\nu(k_1+k_2)^\mu\right)\nonumber\\
&&\nonumber\\
&&\times\left[\delta_{n,|l-m|}-\gamma^5\delta_{n,l+m}\right]\label{eq:56}\\
&&\nonumber\\
&&\nonumber\\
g^l_\alpha(k_1)&\to& g^m_\beta(k_2)+G^n_{\mu\nu}:\nonumber\\
&&\nonumber\\
X^{\mu\nu\alpha\beta}&=&-\frac{i}{2M_4}\left[(-m_lm_m+k_1.k_2)\left(-\eta^{\mu\nu}\eta^{\alpha\beta}+\eta^{\mu\alpha}\eta^{\nu\beta}+\eta^{\mu\beta}\eta^{\nu\alpha}\right)+\eta^{\mu\nu}k_1^\beta k_2^\alpha\right.\nonumber\\
&&\nonumber\\
&&\left.-\left(\eta^{\mu\beta}k_1^\nu
k_2^\alpha+\eta^{\mu\alpha}k_1^\beta k_2^\nu -
\eta^{\alpha\beta}k_1^\mu k_2^\nu+(\nu\leftrightarrow\mu)\right)\right]\delta_{n,|l-m|}\nonumber\\
&&\\
&&\nonumber\\
&&\nonumber\\
q+\bar{q}&\to&g:\nonumber\\
&&\nonumber\\
Z^\mu &=&ig\gamma^\mu t^a\label{eq:58}\\
&&\nonumber\\
&&\nonumber\\
q^l+\bar{q}^m&\to& g^{k\rho}+G^{n\mu\nu}:\nonumber\\
&&\nonumber\\
A^{\mu\nu\rho}&=&\frac{igt^a}{4M_4}\left(-2\eta^{\mu\nu}\gamma^\rho+\eta^{\mu\rho}\gamma^\nu+\eta^{\nu\rho}\gamma^\mu\right)\gamma^5\delta_{l+m,k+n}\label{eq:59}\\
&&\nonumber
\end{eqnarray}

Four diagrams contribute to this process
(Fig.~\ref{fig:7},~\ref{fig:8}). The amplitudes are given by:

\begin{eqnarray}
\mathcal{M}^1&:&\nonumber\\
&&\frac{igt^a}{2M_4}\bar{\nu}(k_2)\left[\gamma^\nu
\epsilon^{*\mu}e^*_{\mu\nu}-\gamma^\nu\epsilon^*_\nu
e^{*\mu}_\mu\right]\gamma^5 u(k_1)\label{eq:60}\\
&&\nonumber\\
\mathcal{M}^2&:&\nonumber\\
&&-\frac{igt^a}{4M_4}\bar{\nu}(k_2)\gamma^\rho\frac{\not{k_1}-\not{p_2}+m_{l-n}}{2(k_1.p_2-m_lm_n)}\left(\eta^{\mu\nu}(\not{p_2}-m_n)+\gamma^\mu k_1^\nu+\gamma^\nu k_1^\mu\right)u(k_1)\epsilon^*_\rho e^*_{\mu\nu}\nonumber\\
&&\label{eq:61}\\
&&\nonumber\\
&&\nonumber\\
\mathcal{M}^3&:&\nonumber\\
&&\frac{igt^a}{4M_4}\bar{\nu}(k_2)\frac{\eta^{\mu\nu}(\not{k_1}-\not{p_1}+m_{l-k})-\gamma^\mu k_2^\nu-\gamma^\nu k_2^\mu}{2(k_1.p_1-m_lm_k)}\gamma^5\left(2k_1^\rho-(\not{p_1}+m_k)\gamma^\rho\right)u(k_1)\epsilon^*_\rho e^*_{\mu\nu}\nonumber\\
&&\label{eq:62}\\
&&\nonumber\\
\mathcal{M}^4&:&\nonumber\\
&&\frac{igt^a}{2M_4}\frac{\bar{\nu}(k_2)}{(k_1+k_2)^2-m_{l+m}^2}\left[(m_km_{k-l-m}+p_1.p_2)\left(\gamma^\rho\eta^{\mu\nu}-\gamma^\mu\eta^{\nu\rho}-\gamma^\nu\eta^{\mu\rho}\right)\right.\nonumber\\
&&\nonumber\\
&&-\left(\eta^{\mu\nu}p_2^\rho-\eta^{\mu\rho}p_1^\nu-\eta^{\nu\rho}p_1^\mu\right)\not{p_1}-2\gamma^\rho
p_1^\mu p_1^\nu+(\gamma^\mu p_1^\nu +\gamma^\nu
p_1^\mu)p_2^\rho{\Huge{]}}u(k_1)\epsilon^*_\rho
e^*_{\mu\nu}\nonumber\\
&&\label{eq:63}
\end{eqnarray}

In the center-of-mass frame, the cross-section is given by
\cite{Peskin:1995ev}:

\begin{equation}
\left(\frac{d\sigma}{d\Omega}\right)_{CM}=\frac{1}{2E_12E_2|\nu_1-\nu_2|}\frac{|{\bf
p_2}|}{(2\pi)^24E_{CM}}|\mathcal{M}(k_1,k_2\to
p_1,p_2)|^2.\label{eq:44}
\end{equation}
Here, $E_1,E_2$ and $\nu_1, \nu_2$ label the initial particle
energies and velocities, and ${\bf p_2}$ is the three-momentum of
the graviton. Since we are interested in the relativistic limit, we
can use the more general form for all four particles with $m=0$:

\begin{equation}
\left(\frac{d\sigma}{d\Omega}\right)_{CM}=\frac{|\mathcal{M}|^2}{32\pi^2|\nu_1-\nu_2|E^2_{CM}}.\label{eq:45}
\end{equation}
Averaging over initial spins and colors and summing over final
polarizations, $|\mathcal{M}|^2$ was calculated using the {\bf
Tensorial 3.0} package for \textit{Mathematica} \cite{Math}. The
cross-section is then given by:
\begin{equation}
<\sigma\nu>=\frac{\alpha_3}{4\pi
M^2_4}d(G)C(r)\pi\frac{1184+105\pi}{1296}.\label{eq:46}
\end{equation}

The fraction of strongly interacting degrees of freedom per KK level
is computed by counting the different allowed interactions. Four
strong interactions can produce gravitons:
\begin{eqnarray}
q+\bar{q}\to g +G\nonumber\\
g+q\to q+G\nonumber\\
g+\bar{q}\to \bar{q}+G\nonumber\\
g+g\to g+G\nonumber
\end{eqnarray}
There are $2\times3\times24$ degrees of freedom for the quarks and
$3\times8$ for the gluons. The fraction can then be calculated:
\begin{equation}
p_{s^2}=24\left[\left(\frac{7}{8}\right)^2\left(\frac{6}{197.5}\right)^2+2\left(\frac{7}{8}\right)\left(\frac{24}{197.5}\right)\left(\frac{6}{197.5}\right)\right]+\left(\frac{24}{197.5}\right)^2\sim
0.19.\label{eq:47}
\end{equation}

 Assuming that all the other graviton production processes
have a cross-section similar to (\ref{eq:46}), $C$ used to
parameterize the graviton production cross-section is given by:

\begin{eqnarray}
C&=&p_{s^2}d(G)C(r)\pi\frac{1184+105\pi}{1296}=2.79,\nonumber\\
&\sim &\mathcal{O}(1).\label{eq:48}
\end{eqnarray}

%\newpage

\renewcommand{\theequation}{B.\arabic{equation}}
% redefine the command that creates the equation no.
\setcounter{equation}{0}  % reset counter
\section{Decay Lifetimes}
\label{appendixB}

The graviton polarization sum is given by \cite{Feng:2003nr}:

\begin{eqnarray}
\sum_s
e^s_{\mu\nu}e^{s*}_{\rho\sigma}&=&B_{\mu\nu\rho\sigma}\nonumber\\
B_{\mu\nu\rho\sigma}(k)&=&2\left(\eta_{\mu\rho}-\frac{k_\mu
k_\rho}{m_n^2}\right)\left(\eta_{\nu\sigma}-\frac{k_\nu
k_\sigma}{m_n^2}\right)+2\left(\eta_{\mu\sigma}-\frac{k_\mu
k_\sigma}{m_n^2}\right)\left(\eta_{\nu\rho}-\frac{k_\nu
k_\rho}{m_n^2}\right)\nonumber\\
&&-\frac{4}{3}\left(\eta_{\mu\nu}-\frac{k_\mu
k_\nu}{m_n^2}\right)\left(\eta_{\rho\sigma}-\frac{k_\rho
k_\sigma}{m_n^2}\right)\label{eq:67}
\end{eqnarray}

The decay width of $G^n(p)\to f^l(k_1)+\bar{f}^m(k_2)$ summed over
final and averaged over initial polarizations is:

\begin{eqnarray}
\Gamma(G^n\to
f^l+\bar{f}^m)&=&\frac{|\textbf{k}_1|}{8\pi m_n^2}\frac{1}{5}\sum_{s,s'}\bar{u}^s(k_2)Y^{\mu\nu}\nu^{s'}(k_1)\bar{\nu}^{s'}(k_1)Y^{*\alpha\beta}u^{s}(k_2)B_{\mu\nu\alpha\beta}\nonumber\\
&=&\frac{1}{160\pi}\frac{m_n^3}{
M_4^2}\left[1-\frac{(m_l-m_m)^2}{m_n^2}\right]^{3/2}\left[1-\frac{(m_l+m_m)^2}{m_n^2}\right]^{5/2}\nonumber\\
&&\times\left[1+\frac{2}{3}\frac{(m_l-m_m)^2}{m_n^2}\right].
\end{eqnarray}
This is suppressed for $l\neq m$. Therefore assuming that the main
decay is into $l=m=n/2$:

\begin{equation}
\Gamma(G^n\to
f^{m=n/2}+\bar{f}^{m=n/2})=\frac{1}{160\pi}\frac{m_n^3}{M_4^2}\left[1-\frac{4m_m^2}{m_n^2}\right]^{5/2}.
\end{equation}

The decay width of $G^n(p)\to B^l(k_1)+B^m(k_2)$ summed over final
and averaged over initial polarizations is given by:

\begin{eqnarray}
\Gamma(G^n\to
B^l+B^m)&=&\frac{|\textbf{k}_1|}{8\pi m_n^2}\frac{1}{5}X^{\mu\nu\alpha\beta}X^{*\rho\sigma\lambda\kappa}\left(\eta_{\alpha\lambda}-\frac{k_{1\alpha}k_{1\lambda}}{m_m^2}\right)\left(\eta_{\beta\kappa}-\frac{k_{2\beta}k_{2\kappa}}{m_l^2}\right)B_{\mu\nu\rho\sigma}\nonumber\\
&=&\frac{\cos^2\theta_W}{80\pi}\frac{m_n^3}{M_4^2}\left[1-\frac{\left(m_l-m_m\right)^2}{m_n^2}\right]^{5/2}\left[1-\frac{\left(m_l+m_m\right)^2}{m_n^2}\right]^{1/2}\nonumber\\
&&\times\left[\frac{13}{6}+\frac{7}{3}\frac{\left(m_l+m_m\right)^2}{m_n^2}+\frac{1}{2}\frac{\left(m_l+m_m\right)^4}{m_n^4}\right].\nonumber\\
&&\label{eq:68}
\end{eqnarray}
We see again that the decay into KK modes other than $n/2$ is
suppressed. Therefore for all $n>1$, assuming now that the main
decay is in fact into $B^{n/2}+B^{n/2}$ with an extra factor of
$1/2$ for identical particles in the final state:

\begin{equation}
\Gamma(G^n\to
2B^{m=n/2})=\frac{\cos^2\theta_W}{80\pi}\frac{m_n^3}{M_4^2}\left[1-\frac{4m_m^2}{m_n^2}\right]^{1/2}\left[\frac{13}{12}+\frac{14}{3}\frac{m_m^2}{m_n^2}+4\frac{m_m^4}{m_n^4}\right],\label{eq:49}
\end{equation}
with $\Delta_n \equiv m_n-2m_m\sim m_{G^n}-m_{B^n}<< m_n$:
\begin{equation}
\Gamma\sim\frac{\sqrt{2}\cos^2\theta_W}{32\pi}\frac{m_n^3}{M_4^2}\sqrt{\frac{\Delta_n}{m_n}}.\label{eq:50}
\end{equation}

For any $n$ decay into a KK gauge boson and a photon:
\begin{equation}
\Gamma(G^n\to
B^n+\gamma)=\frac{\cos^2\theta_W}{30\pi}\frac{\Delta_n^3}{M_4^2}\left[6+3\frac{m_{B^n}^2}{m_{G^n}^2}+\frac{m_{B^n}^4}{m_{G^n}^4}\right].\label{eq:51}
\end{equation}

 \end{document}